\documentclass[epj]{svjour}                             
\usepackage{graphics}
\begin{document}
\title{Microscopic approach to the nucleon-nucleon effective interaction                 
and nucleon-nucleon scattering 
in symmetric and isospin-asymmetric nuclear matter 
}
\author{F. Sammarruca}                       
\institute{Physics Department, University of Idaho, Moscow, ID 83844-0903, U.S.A}
\date{}

\abstract{ 
After reviewing our microscopic approach to nuclear and neutron-rich matter, 
we focus on how nucleon-nucleon scattering is impacted by the presence of 
a dense hadronic medium, with special emphasis on the case where neutron and proton
densities are different. 
We discuss in detail medium and isospin asymmetry effects on the total elastic cross section
and the mean free path of a neutron or a proton in isospin asymmetric nuclear matter. 
We point out that in-medium cross sections play an important role in heavy-ion simulations
aimed at extracting constraints on the symmetry potential. 
 We argue that medium and isospin dependence of microscopic cross sections are the result of 
a complex balance among various effects, and cannot be simulated with 
a simple phenomenological model. 
}
\titlerunning{Microscopic approach to the nucleon-nucleon effective...}                                      
\maketitle

\section{Introduction}
\label{intro} 
                            
In this article, we will be concerned with 
hadronic interactions
in the nuclear medium, an                                                                 
 issue  which goes to the very core 
of nuclear physics. In fact, our present knowledge of the nuclear force in free space is, in itself, 
the result of decades of struggle \cite{Mac89} which will not be reviewed here.              
The nature of the nuclear force in the medium is of course an even more complex problem, 
as it involves aspects of the force that cannot be constrained
through free-space nucleon-nucleon (NN) scattering. Predictions of properties of nuclei are 
the ultimate test for many-body theories. 

Nuclear matter is an alternative and convenient theoretical laboratory to test many-body theories. By ``nuclear matter" we mean an infinite system 
of nucleons acted on by their mutual strong forces and no electromagnetic interactions. Nuclear matter 
is characterized by its energy/particle as a function of density and other thermodynamic quantities, as  
appropriate (e.g.~temperature). Such relation is known as the nuclear matter equation of state (EoS). 
The translational invariance of the system facilitates theoretical calculations. At the same time, adopting     
what is known as 
the ``local density approximation", one can use the EoS to obtain information on finite systems. This procedure 
is applied,   
for instance, in Thomas-Fermi calculations within the liquid drop model, where an appropriate energy functional is 
written in terms of the EoS \cite{Oya98,Furn,SL09}.

Isospin-asymmetric nuclear matter (IANM) simulates the interior of a nucleus with unequal densities of protons and neutrons.
The equation of state of (cold) IANM is then a function of density as well as the relative concentrations 
of protons and neutrons. 

The recent and fast-growing interest in IANM stems from its close connection to the physics of neutron-rich nuclei, or,
more generally, isospin-asymmetric nuclei, including the very ``exotic" ones known as ``halo" nuclei. 
At this time, the boundaries of the nuclear chart are uncertain, with a few hundred stable nuclides 
known to exist and a few thousand believed to exist. 
The future Facility for Rare Isotope Beams (FRIB) is expected                                                
to deliver intense beams of rare isotopes, the study of which can provide crucial 
information on short-lived elements normally not found on earth.                  
Thus, this new experimental program will have widespread impact, ranging from the origin of elements to the              
evolution of the cosmos. 
In the meantime, systematic investigations to determine
the properties of asymmetric nuclear matter and to constrain the symmetry energy are proliferating at existing facilities. 

It is the focal point of this article to review and discuss our approach to the 
devolopment of effective NN interactions in IANM, with particular 
emphasis on its applications to in-medium isospin-dependent NN cross sections and related issues.

The article is organized as follows: In Section {\bf 2}, we set the stage by presenting a 
brief overview of facts and phenomenology about IANM. We then proceed to describe our
microscopic approach for calculating the energy/particle in IANM (Section {\bf 3}).          
 Although the EoS {\it per se} is not the focal point of  this article, this step
is important to elucidate how the self-consistent determination of the effective interaction and the 
(isospin-dependent) single-particle potential is obtained. The latter is closely related to the 
nucleon effective masses, which then become a crucial ingredient in the calculation of the 
isospin-dependent effective cross sections. Those will be confronted in Section {\bf 4}. 
In Section {\bf 5} we will discuss the mean free path of nucleons in nuclear matter and its 
relation to the effective cross sections. 
A brief summary and conclusive remarks are contained in Section {\bf 6}. 

\section{Facts about isospin-asymmetric nuclear matter} 
\label{sec2} 

Asymmetric nuclear matter is characterized by the neutron density, 
$\rho_n$, and the proton density, $\rho_p$.                                                                   
In infinite matter, they are obtained by summing the neutron or proton states per volume (up to their respective 
Fermi momenta, $k^{n}_{F}$ or $k^{p}_{F}$) and applying the appropriate degeneracy factor. The result is 
\begin{equation}
  \rho_i =\frac{ (k^{i}_{F})^3}{3 \pi ^2} ,   \label{rhonp}   
\end{equation}
with $i=n$ or $p$. 

It is more convenient to refer to the total density
$\rho = \rho_n + \rho_p$ and the asymmetry (or neutron excess) parameter
$\alpha = \frac{ \rho_n - \rho_p}{\rho}$. 
Clearly, $\alpha$=0 corresponds to symmetric matter and 
$\alpha$=1 to neutron matter.                       
In terms of $\alpha$ and the average Fermi momentum, $k_F$, related to the total density in the usual way, 
\begin{equation}
  \rho =\frac{2 k_F^3}{3 \pi ^2} ,   \label{rho}   
\end{equation}
the neutron and proton Fermi momenta can be expressed as 
\begin{equation}
 k^{n}_{F} = k_F{(1 + \alpha)}^{1/3}            \label{kfn}
\end{equation}
and 
\begin{equation}
 k^{p}_{F} = k_F{(1 - \alpha)}^{1/3} ,            \label{kfp} 
\end{equation}
 respectively.

Expanding 
the energy/particle in IANM  with respect to the asymmetry parameter yields
\begin{equation}
e(\rho, \alpha) = e_0({\rho}) + \frac{1}{2} \Big (\frac{\partial ^2 e(\rho,\alpha)}{\partial \alpha ^2}\Big )_{\alpha=0}\alpha ^2 +{\cal O}(\alpha ^4) \; , \label{exp}  
\end{equation}
where the first term is the energy per particle in symmetric matter and 
the coefficient of the quadratic term is identified with the symmetry energy,
$e_{sym}$. In the Bethe-Weizs{\" a}cker formula for the nuclear binding energy, it represents the amount of binding a nucleus has 
to lose when the numbers of protons and neutrons are unequal.                                             
The symmetry energy is also closely related to 
the neutron $\beta$-decay in dense matter, whose threshold depends on the proton fraction. 
A typical value for $e_{sym}$               
at nuclear matter density ($\rho_0$) is 30 MeV, 
with theoretical predictions spreading approximately between 26 and 35 MeV.

To a very good degree of approximation, 
the energy per particle in IANM can be written as 
\begin{equation}
e(\rho, \alpha) \approx e_0({\rho}) + e_{sym}(\rho)\alpha ^2.   \label{e}                    
\end{equation} 
The effect of a term of fourth degree in the asymmetry parameter (${\cal O}(\alpha ^4)$) on the bulk properties of neutron stars 
is small, although it may impact the proton fraction at high density. 
More generally, 
non-quadratic terms are usually associated with isovector pairing, which is a surface effect and thus vanishes
in infinite matter \cite{Steiner}. 

Equation~(\ref{e}) displays a convenient separation between the symmetric and the aymmetric parts of the EoS, 
which facilitates the identification of observables that may be sensitive, for instance, mainly to the 
symmetry energy.                                                                              
Typically, constraints are extracted from heavy-ion collision simulations 
based on transport models. 
Isospin diffusion and the ratio of neutron and proton spectra are among the 
observables used in these analyses. 
 For a recent review on available constraints the reader is referred to Ref.~\cite{Tsang+}. 

Empirical investigations appear to agree reasonably well on the following parametrization 
of the symmetry energy: 
\begin{equation}
e_{sym}(\rho) = 12.5 \, MeV \Big (\frac{\rho}{\rho_0}\Big )^{2/3} +                       
17.5 \, MeV \Big (\frac{\rho}{\rho_0}\Big )^{\gamma_i} \; ,               \label{es} 
\end{equation} 
where $\rho_0$ is the saturation density.                                                             
The first term is the kinetic contribution and 
 $\gamma_i$ (the exponent appearing in the potential energy part) is found to be between 0.4 and 1.0. 
Recent measurements of elliptic flows in $^{197}$Au + $^{197}$Au reactions at GSI  at  
400-800 MeV/nucleon 
favor a potential energy term with $\gamma_i$ equal to 0.9 $\pm$ 0.4.           
Giant dipole resonance excitation in fusion reactions \cite{GDR} is also sensitive to
the symmetry energy, since the latter is responsible for isospin equilibration 
in isospin-asymmetric collisions.

Isospin-sensitive observables can also be identified among the properties of normal nuclei. 
The neutron skin of neutron-rich nuclei is a powerful isovector observable, being sensitive to the   
slope of the symmetry energy, which determines to which extent neutrons will tend to spread outwards 
 to form the skin. 
                           
Parity-violating electron scattering experiments are now a realistic option        
to determine neutron distributions with unprecedented accuracy. The neutron radius of 
$^{208}$Pb is expected to be re-measured                                                        
at the Jefferson Laboratory in the PREXII experiment planned for the near future.                        
Parity-violating electron scattering at low momentum transfer is especially suitable to probe neutron densities, as the 
 $Z^0$ boson couples primarily to neutrons. A much higher level of accuracy can be achieved with 
electroweak probes than with hadronic scattering.
With the success of this program, 
 reliable empirical information on neutron skins will be able to provide, in turn, more stringent constraint on the 
density dependence of the symmetry energy.

A measure for the density dependence of the symmetry energy is 
 the parameter defined as 
\begin{equation}
L = 3 \rho_0 \Big (\frac{\partial e_{sym}(\rho)}{\partial \rho}\Big )_{\rho_0} \approx 
 3 \rho_0 \Big (\frac{\partial e_{n.m.}(\rho)}{\partial \rho}\Big )_{\rho_0} \, , 
\label{L} 
\end{equation} 
where we have used Eq.~(\ref{e}) with $\alpha$=1. 
Thus, $L$ is sensitive to the gradient of the energy per particle in neutron matter ($e_{n.m.}$), that is, the 
neutron matter pressure. 
As to be expected on physical grounds, the neutron skin, given by                                  
\begin{equation}
S = \sqrt{<r_n^2>} - \sqrt{<r_p^2>} \, \, , 
\label{S} 
\end{equation} 
is highly sensitive to the same energy gradient.

Predictions of $L$ by                      
phenomenological models show a very large spreading. Values ranging from 
-50 to +100 MeV are found from the numerous
parametrizations of Skyrme interactions (see Ref.~\cite{BA05} and references therein),               
 all chosen to fit the binding energies and the 
charge radii of a large number of nuclei.  

In Ref.~\cite{Tsang+},                 
values for the symmetry energy and the $L$ parameter centered around 32.5 MeV 
and 70 MeV, respectively, are obtained, both from nuclear structure and heavy-ion collision
measurements, for densities ranging between 
0.3$\rho_0$ and                                                          
$\rho_0$.

Typically, parametrizations like the one given in Eq.~(\ref{es}) are valid 
at or below the saturation density. Efforts to constrain the behavior of the symmetry energy
at higher densities 
are being pursued through observables such as $\pi ^-/\pi^+$ ratio, 
$K ^+/K^0$ ratio, neutron/proton differential transverse flow, or nucleon elliptic flow \cite{Ko09}. 

\begin{figure}[!t] 
\centering         
\vspace*{0.17cm}
\hspace*{-0.5cm}
\scalebox{0.6}{\includegraphics{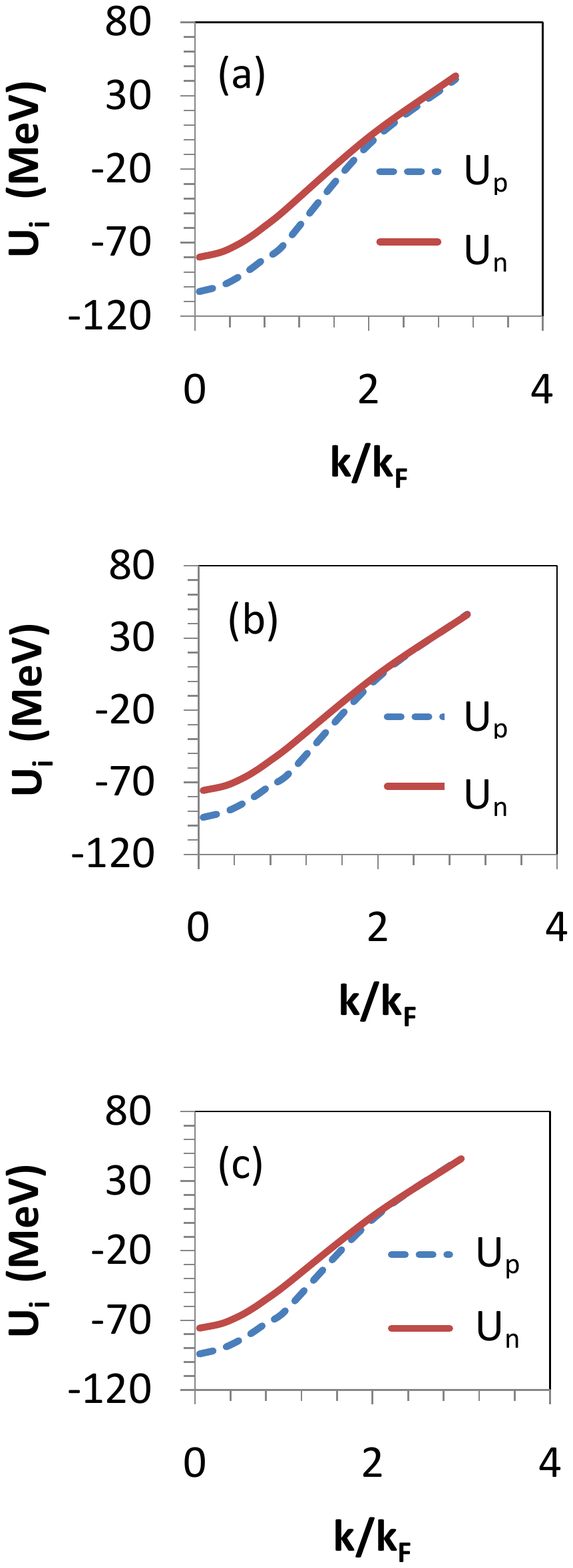}}
\vspace*{-1.9cm}
\caption{(Color online) Momentum dependence of the single-nucleon potential in IANM, $U_i$ $(i=n,p)$, 
predicted with Bonn A (a), Bonn B (b), and Bonn C (c). 
The total density is equal to 0.185 fm$^{-3}$, and the isospin asymmetry parameter is equal to
0.4. The momentum is given in units of the (average) Fermi momentum, which is equal to 1.4 fm$^{-1}$. 
} 
\label{unp}
\end{figure}

Another important quantity which emerges from studies of IANM is the symmetry potential.
Its definition stems from the observation that the single-particle potentials 
experienced by the proton and the neutron in IANM, $U_{n/p}$, are different from each other and satisfy      
the approximate relation 
\begin{equation}
U_{n/p}(k,\rho,\alpha) \approx U_{n/p}(k,\rho,\alpha=0) \pm U_{sym}(k,\rho)\;\alpha \; , 
\label{Unp}
\end{equation}
where the +(-) sign refers to neutrons (protons), and               
\begin{equation}
U_{sym}=\frac{U_{n} - U_p}{2\alpha} \; .                  
\label{Usym} 
\end{equation}
Thus, 
one can expect isospin splitting of the single-particle potential to be effective in separating           
the collision dynamics of neutrons and protons. 
In a neutron-rich environment, the symmetry potential tends to expel neutrons and attract protons, 
thus providing the opportunity of detecting sensitivity to the symmetry energy in observables such as the yield ratios of ejected neutrons/protons or the rate of isospin 
diffusion \cite{Tsang+}.                               
The splitting of the single-nucleon potentials in IANM as a function of the momentum
is shown in Fig.~\ref{unp} for three different meson-theoretic potentials \cite{Mac89}. 

Furthermore, $U_{sym}$, being proportional to the gradient between the single-neutron and
the single-proton potentials, 
should be comparable with the Lane potential \cite{Lane}, namely the isovector 
 part of the nuclear optical potential. Optical potential analyses (in isospin unsaturated nuclei) can then 
help constrain this quantity and, in turn, the symmetry energy.           

\section{Our microscopic approach to isospin-asymmetric nuclear matter}
\label{sec3}

\subsection{The two-body sector} 
\label{subsec3a}

Our approach is
{\it ab initio} in that the starting point of the many-body calculation is a realistic NN interaction which is then applied in the 
nuclear medium without any additional free parameters. 
Thus the first question to be confronted concerns the choice of the ``best" NN interaction. 
After the development of Quantum Chromodynamics (QCD) and the understanding of its symmetries,  
chiral effective theories \cite{chi} were developed as a way to respect the 
symmetries of QCD while keeping the degrees of freedom (nucleons and pions) typical of low-energy nuclear physics. However, 
chiral perturbation theory (ChPT)
has definite limitations as far as the range of allowed momenta is concerned. 
For the purpose of applications in dense matter, where higher and higher momenta become involved     
with increasing Fermi momentum, NN potentials based on ChPT are unsuitable.       

Relativistic meson theory is an appropriate framework to deal with the high momenta encountered in dense
matter. In particular, 
the one-boson-exchange (OBE) model has proven very successful in describing NN elastic data in free space 
up to high energies 
and has a good theoretical foundation. 
Among the many available OBE potentials, some being part of the ``high-precision generation" \cite{pot1,pot2,pot3}, 
we seek a momentum-space potential developed within a relativistic scattering equation, such as the 
one obtained through the Thompson \cite{Thom} three-dimensional reduction of the Bethe-Salpeter equation \cite{BS}. 
Furthermore, we require a potential that uses 
the pseudovector coupling for the interaction of nucleons with pseudoscalar mesons. 
With these constraints in mind, 
as well as the requirement of a good description of the NN data, 
Bonn B \cite{Mac89} is a reasonable choice. As is well known, the NN potential model dependence
of nuclear matter predictions is not negligible. The saturation points obtained with different NN potentials
move along the famous ``Coester band" depending on the strength of the tensor force, with the weakest tensor
force yielding the largest attraction. This can be understood in terms of medium effects (particularly 
Pauli blocking) reducing the (attractive) second-order term in the expansion of the reaction matrix. 
A large second-order term will undergo a large reduction in the medium. Therefore, noticing that the second-order term
is dominated by the tensor component of  the force, nuclear potentials with a strong tensor component will
yield less attraction in the medium. 
For the same reason (that is, the role of the tensor force in  
nuclear matter), 
the potential model dependence is strongly reduced in pure (or nearly pure) neutron matter, due to the  
absence of isospin-zero partial waves.

\begin{figure}[!t] 
\centering         
\vspace*{-3.2cm}
\hspace*{-2.5cm}
\scalebox{0.9}{\includegraphics{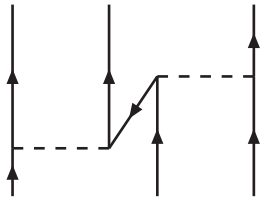}}
\vspace*{-19.0cm}
\caption{Three-body force due to virtual pair excitation.                                          
} 
\label{3b}
\end{figure}

In closing this section, we wish to highlight          
the most important aspect of the {\it ab initio} approach: namely, the only free parameters of the
model (the parameters of the NN potential)                                               
are determined by fitting the free-space NN data and never readjusted in the medium. In other
words, the model parameters are tightly constrained and the calculation in the medium is 
parameter free. 
The presence of free parameters in the medium would generate effects and sensitivities which are hard to
control and reduce the predictive power of the theory. 

\subsection{The Dirac-Brueckner-Hartree-Fock approach to symmetric and  asymmetric nuclear matter}
\label{subsec3b} 

The main strength of the DBHF approach is its inherent ability to account for important three-body forces   
through its density dependence. 
In Fig.~\ref{3b} we show a three-body force (TBF) originating from virtual excitation of a nucleon-antinucleon pair, 
known as ``Z-diagram".                                                                                              
The main feature of                                     
the DBHF method turns out to be closely related to 
the TBF depicted in Fig.~\ref{3b}, as we will argue next. In the DBHF approach, one describes the positive energy solutions
of the Dirac equation in the medium as 
\begin{equation}
u^*(p,\lambda) = \left (\frac{E^*_p+m^*}{2m^*}\right )^{1/2}
\left( \begin{array}{c}
 {\bf 1} \\
\frac{\sigma \cdot \vec {p}}{E^*_p+m^*} 
\end{array} 
\right) \;
\chi_{\lambda},
\label{ustar}
\end{equation}
where the nucleon effective mass, $m^*$, is defined as $m^* = m+U_S$, with $U_S$ an attractive scalar potential.
(This will be derived below.) 
It can be shown that both the description of a single-nucleon via Eq.~(\ref{ustar}) and the evaluation of the 
Z-diagram, Fig.~\ref{3b}, generate a repulsive effect on the energy per particle in symmetric nuclear matter which depends on the density approximately
as 
\begin{equation}
\Delta E \propto  \left (\frac{\rho}{\rho_0}\right )^{8/3} \, , 
\label{delE} 
\end {equation}
and provides the saturating mechanism missing from conventional Brueckner-Hartree-Fock (BHF) calculations. 
(Alternatively, explicit TBF are used along with the BHF method in order to achieve a similar result.) 
Brown showed that the bulk of the desired effect can be obtained as a lowest order (in $p^2/m$) relativistic correction
to the single-particle propagation \cite{GB87}. 
With the in-medium spinor as in Eq.~(\ref{ustar}), the correction to the free-space spinor can be written 
approximately as 
\begin{equation}
u^*(p,\lambda) -u(p,\lambda)\approx                                                  
\left( \begin{array}{c}
 {\bf 0} \\
-\frac{\sigma \cdot \vec {p}}{2 m^2}U_S
\end{array} 
\right) \;
\chi_{\lambda},
\label{delu} 
\end{equation}
where for simplicity the spinor normalization factor has been set equal to 1, in which case it 
is clearly seen that the entire effect originates from the modification of the spinor's lower component. 
By expanding the single-particle energy to order $U_S^2$, Brown showed that the correction to the 
energy consistent with Eq.~(\ref{delu}) can be written as $\frac{p^2}{2m}(\frac{U_S}{m})^2$. He then proceeded to 
estimate the correction to the energy per particle and found it to be approximately as given in Eq.~(\ref{delE}).

The approximate equivalence of the effective-mass description of Dirac states and the contribution from the Z-diagram 
has a simple intuitive explanation in the observation 
that Eq.~(\ref{ustar}), like any other solution of the Dirac equation,
can be written as a superposition of positive and negative energy solutions. On the other hand, the ``nucleon" in the 
middle of the Z-diagram, Fig.~\ref{3b}, can be viewed as the superposition of positive and negative energy states. 
In summary, the DBHF method effectively takes into account a particular class of 
TBF, which are crucial for nuclear matter saturation.

Having first summarized the main DBHF philosophy, 
we now proceed to review our DBHF calculation of IANM \cite{AS03,FS10}. 
In the end, this will take us back to the crucial point of the DBHF approximation, Eq.~(\ref{ustar}). 

As mentioned in the previous subsection, 
we start from the Thompson \cite{Thom} relativistic three-dimensional reduction 
of the Bethe-Salpeter equation \cite{BS}. The Thompson equation is applied to nuclear matter in
strict analogy to free-space scattering and reads, in the nuclear matter rest frame,                 
\begin{eqnarray}
&& g_{ij}(\vec q',\vec q,\vec P,(\epsilon ^*_{ij})_0) = v_{ij}^*(\vec q',\vec q) \nonumber \\            
&& + \int \frac{d^3K}{(2\pi)^3}v^*_{ij}(\vec q',\vec K)\frac{m^*_i m^*_j}{E^*_i E^*_j}
\frac{Q_{ij}(\vec K,\vec P)}{(\epsilon ^*_{ij})_0 -\epsilon ^*_{ij}(\vec P,\vec K)}           
\nonumber \\ 
&& 
\times g_{ij}(\vec K,\vec q,\vec P,(\epsilon^*_{ij})_0) \, ,                                   
\label{gij}
\end{eqnarray}                    
where $g_{ij}$ is the in-medium reaction matrix 
($ij$=$nn$, $pp$, or $np$), and the                                      
asterix signifies that medium effects are applied to those quantities. Thus the NN potential, 
$v_{ij}^*$, is constructed in terms of effective Dirac states (in-medium spinors) as explained above. 
In Eq.~(\ref{gij}),                                  
$\vec q$, $\vec q'$, and $\vec K$ are the initial, final, and intermediate
relative momenta, and $E^*_i = \sqrt{(m^*_i)^2 + K^2}$. 
The momenta of the two interacting particles in the nuclear matter rest frame have been expressed in terms of their
relative momentum and the center-of-mass momentum, $\vec P$, through
\begin{equation} 
\vec P = \vec k_{1} + \vec k_{2}       \label{P}    
\end{equation} 
and 
\begin{equation} 
\vec K = \frac{\vec k_{1} - \vec k_{2}}{2} \, .  \label{K}
\end{equation}                    
The energy of the two-particle system is 
\begin{equation} 
\epsilon ^*_{ij}(\vec P, \vec K) = 
e^*_{i}(\vec P, \vec K)+  
e^*_{j}(\vec P, \vec K)   
\label{eij}
\end{equation} 
 and $(\epsilon ^*_{ij})_0$ is the starting energy.
 The single-particle energy $e_i^*$ includes kinetic energy and potential 
 energy contributions (see Eq.~(\ref{spe}) below).                               
The Pauli operator, $Q_{ij}$, prevents scattering to occupied $nn$, $pp$, or $np$ states.            
 To eliminate the angular
dependence from the kernel of Eq.~(\ref{gij}), it is customary to replace the exact
Pauli operator with its angle-average. 
Detailed expressions for the Pauli operator                     
and the average center-of-mass momentum in the case of two different Fermi seas  
can be found in Ref.~\cite{AS03}.                              

With the definitions
\begin{equation} 
G_{ij} = \frac{m^*_i}{E_i^*(\vec{q'})}g_{ij}
 \frac{m^*_j}{E_j^*(\vec{q})}             
\label{Gij}
\end{equation} 
and 
\begin{equation} 
V_{ij}^* = \frac{m^*_i}{E_i^*(\vec{q'})}v_{ij}^*
 \frac{m^*_j}{E_j^*(\vec{q})} \, ,        
\label{Vij}
\end{equation} 
 one can rewrite Eq.~(\ref{gij}) as
\begin{eqnarray}
&& G_{ij}(\vec q',\vec q,\vec P,(\epsilon ^*_{ij})_0) = V_{ij}^*(\vec q',\vec q) \nonumber \\[4pt]
&& + \int \frac{d^3K}{(2\pi)^3}V^*_{ij}(\vec q',\vec K)
\frac{Q_{ij}(\vec K,\vec P)}{(\epsilon ^*_{ij})_0 -\epsilon ^*_{ij}(\vec P,\vec K)} 
\nonumber \\ 
&&  \times G_{ij}(\vec K,\vec q,\vec P,(\epsilon^*_{ij})_0) \, ,                                    
\label{Geq}
\end{eqnarray}                    
which is our working equation and has the convenient feature of
being formally identical to its non-relativistic counterpart.

The goal is to determine self-consistently the nuclear matter single-particle potential   
which, in IANM, will be different for neutrons and protons. 
To facilitate the description of the procedure, we will use a schematic
notation for the neutron/proton potential.                                                   
We write, for neutrons,
\begin{equation}
U_n = U_{np} + U_{nn} \; , 
\label{un}
\end{equation}
and for protons
\begin{equation}
U_p = U_{pn} + U_{pp} \, , 
\label{up}
\end{equation}
where each of the four pieces on the right-hand-side of Eqs.~(\ref{un}-\ref{up}) signifies an integral of the appropriate 
$G$-matrix elements ($nn$, $pp$, or $np$) obtained from Eq.~(\ref{Geq}).                                           
Clearly, the two equations above are coupled through 
the $np$ component and so they must be solved simultaneously. Furthermore, 
the $G$-matrix equation and Eqs.~(\ref{un}-\ref{up})  
are coupled through the single-particle energy (which includes the single-particle
potential, itself defined in terms of the $G$-matrix). So we have a coupled system to be solved self-consistently.

Before proceeding with the self-consistency, 
one needs an {\it ansatz} for the single-particle potential. The latter is suggested by 
the most general structure of the nucleon self-energy operator consistent with 
all symmetry requirements. That is: 
\begin{equation}
{\cal U}_i({\vec p}) =  U_{S,i}(p) + \gamma_0  U_{V,i}^{0}(p) - {\vec \gamma}\cdot {\vec p}  U_{V,i}(p) \, , 
\label{Ui1}
\end{equation}
where $U_{S,i}$ and 
$U_{V,i}$ are an attractive scalar field and a repulsive vector field, respectively, with 
$ U_{V,i}^{0}$ the timelike component of the vector field. These fields are in general density and momentum dependent. 
We take             
\begin{equation}
{\cal U}_i({\vec p}) \approx U_{S,i}(p) + \gamma_0 U_{V,i}^{0}(p) \, ,                                            
\label{Ui2}
\end{equation}
which amounts to assuming that the spacelike component of the vector field is much smaller than 
 both $U_{S,i}$ and $U_{V,i}^0$. Furthermore, neglecting the momentum dependence of the scalar and
vector fields and inserting Eq.~(\ref{Ui2}) in the Dirac equation for neutrons/protons propagating in 
nuclear matter,
\begin{equation}
(\gamma _{\mu}p^{\mu} - m_i - {\cal U}_i({\vec p})) u^*_i({\vec p},\lambda) = 0  \, ,                                                       
\label{Dirac1} 
\end{equation}
naturally leads to rewriting the Dirac equation in the form 
\begin{equation}
(\gamma _{\mu}(p^{\mu})^* - m_i^*) u^*_i({\vec p},\lambda) = 0  \, ,                                                    
\label{Dirac2} 
\end{equation}
with positive energy solutions as in Eq.~(\ref{ustar}), $m_i^* = m + U_{S,i}$, and 
\begin{equation}
(p^0)^* = p^0 - U_{V,i}^0 (p) \, .                                                                 
\label{p0}
\end{equation}
The subscript ``$i$'' signifies that these parameters are different for protons and
neutrons. 

As in the symmetric matter case \cite{BM84}, evaluating  the expectation value of Eq.~(\ref{Ui2})       
leads to a parametrization of 
the single particle potential for protons and neutrons (Eqs.(\ref{un}-\ref{up})) in terms of the 
constants $U_{S,i}$ and $U_{V,i}^0$ which is given by      
\begin{equation}
U_i(p) = \frac{m^*_i}{E^*_i}<{\vec p}|{\cal U}_i({\vec p})|{\vec p}> = 
\frac{m^*_i}{E^*_i}U_{S,i} + U_{V,i}^0 \; .      
\label{Ui3}
\end{equation}
These are the single-nucleon potentials displayed in Fig.~1. 
Also, 
\begin{equation}
U_i(p) =                                                              
\sum_{j=n,p} 
\sum_{p' \le k_F^j} G_{ij}({\vec p},{\vec p}') \; , 
\label{Ui4}
\end{equation}
which, along with Eq.~(\ref{Ui3}), allows the self-consistent determination of the single-particle
potentials                     
displayed in Fig.~\ref{unp}. 

From the Dirac equation, Eq.~(\ref{Dirac1}), 
the kinetic contribution to the single-particle energy is
\begin{equation}
T_i(p) = \frac{m^*_i}{E^*_i}<{\vec p}|{\vec \gamma} \cdot {\vec p} + m|{\vec p}> =     
\frac{m_i m^*_i + {\vec p}^2}{E^*_i} \; , 
\label{KE}    
\end{equation}
and the single-particle energy is 
\begin{equation}
e^*_i(p) = T_i(p) + U_i(p) = E^*_i + U^0_{V,i} \; . 
\label{spe}
\end{equation}
The constants $m_i^*$ and 
\begin{equation}
U_{0,i} = U_{S,i} + U_{V,i}^0      
\label{U0i} 
\end{equation}
are convenient to work with as they 
facilitate          
the connection with the usual non-relativistic framework \cite{HT70}.                       

Starting from some initial values of $m^*_i$ and $U_{0,i}$, the $G$-matrix equation is 
 solved and a first approximation for $U_{i}(p)$ is obtained by integrating the $G$-matrix 
over the appropriate Fermi sea, see Eq.~(\ref{Ui4}). This solution is 
again parametrized in terms of a new set of constants, determined by fitting the parametrized $U_i$, 
Eq.~(\ref{Ui3}), 
to its values calculated at two momenta, a procedure known as the ``reference spectrum approximation". 
The iterative procedure is repeated until satisfactory convergence is reached.     

Finally, the energy per neutron or proton in nuclear matter is calculated from 
the average values of the kinetic and potential energies as 
\begin{equation}
\bar{e}_{i} = \frac{1}{A}<T_{i}> + \frac{1}{2A}<U_{i}> -m \; . 
\label{ei}
\end{equation}
 The EoS, or energy per nucleon as a function of density, is then written as
\begin{equation}
    \bar{e}(\rho_n,\rho_p) = \frac{\rho_n \bar{e}_n + \rho_p \bar{e}_p}{\rho} \, , 
\label{enp} 
\end{equation}
or 
\begin{equation}
    \bar{e}(k_F,\alpha) = \frac{(1 + \alpha) \bar{e}_n + (1-\alpha) \bar{e}_p}{2} \, . 
\label{eav} 
\end{equation}
Clearly, symmetric nuclear matter is obtained as a by-product of the calculation described above 
by setting $\alpha$=0, whereas $\alpha$=1 corresponds to pure neutron matter. 

\section{Medium- and isospin-dependent NN cross sections} 
\label{sec4}

\subsection{General aspects} 
\label{subsec4a}

 Transport equations                                                            
 describe the evolution of a \\ non-equilibrium gas of 
 strongly interacting hadrons. In models based on 
 the Boltzmann-Uehling-Uhlenbeck (BUU) equation \cite{BBU1,BBU2}, 
 particles drift in the presence of the mean field while undergoing    
 two-body collisions, which require the knowledge of in-medium two-body cross sections.
 In a microscopic approach, 
 both the mean field and the binary collisions are calculated 
  self-consistently starting from the bare two-nucleon force.

 We will present microscopic predictions of NN total elastic cross sections in isospin symmetric and 
 asymmetric nuclear matter. In asymmetric matter, the cross section becomes
 isospin dependent beyond the usual and well-known differences between the $np$ and the $pp/nn$ 
 cases. Here, we are referring to isospin dependence induced by medium asymmetries, meaning that, even in the same 
isospin state, the $nn$, $pp$, and $np$ interactions are different because of different 
 relative proton and neutron concentrations. Also, we are only                    
 concerned with  
the strong interaction contribution to the cross section
 (Coulomb effects on the
 $pp$ cross section or charge-symmetry and charge-independence breaking effects are not 
considered.) 

As mentioned in the Introduction, 
there is increasing interest in studying      
isospin asymmetries in nuclear matter.     
 Collisions of neutron-rich nuclei 
 are capable of 
producing extended regions
of space/time where both the total nucleon density and the neutron/proton 
asymmetry are large.                                                               
Isospin-dependent BUU transport models                       
include isospin-sensitive collision dynamics through the elementary $pp$, $nn$, and $np$
cross sections and the mean field, which is different for protons and        
neutrons.    

The in-medium cross sections are driven by the scattering amplitude and also by kinematic factors, i.~e., 
entrance flow and density of states in the exit channel, both of which are related to the effective mass
(and thus, the nucleon self-energy). 
In-medium cross sections depend non-trivially on several variables, such as the relative momentum 
of the nucleon pair, the total momentum of the pair in the nuclear matter rest frame, and,      
 in the case of asymmetric matter, two 
different densities.                                       
To facilitate applications in reactions, 
these multiple dependences have been handled in different ways and with different levels 
of approximations, with the result that predictions can be quite different from one another.
Model differences include, for instance, whether or not medium effects are present both in the $G$-matrix and          
the density of states; or whether Pauli blocking effects are taken into account in both the final and intermediate configurations.  

 In a simpler approach, the assumption is made that the transition matrix
 in the medium is approximately the same as the one in vacuum and that         
 medium effects on the cross section 
 come in only through the use of  nucleon effective masses in the phase space factors 
 \cite{PP,Gale,LC}. Concerning microscopic calculations,  
  some can be found, for instance, in                          
 Refs.~\cite{LM,Fuchs,SSRL}, but considerations of medium asymmetries are 
 not included in those predictions. 
In Ref.~\cite{Zhang}, the Brueckner-Hartree-Fock method with the Argonn $v_{14}$ potential including 
the contribution of microscopic three-body forces is employed. 
A brief review of empirical and theoretical findings is given next.

In-medium cross sections can provide information 
on the mean free path of nucleons in nuclear matter and thus nuclear 
transparency. The latter is obviously related to the total reaction 
cross section of a nucleus, which, in turn, can be used to extract
 nuclear r.m.s. radii within Glauber-type models \cite{Glauber}. Therefore,  accurate in-medium 
 {\it isospin-dependent} NN cross sections can ultimately be very valuable to obtain 
 information about the size of exotic, neutron-rich nuclei. 
 In summary, it is important to 
 investigate to which extent the in-medium
 cross sections are sensitive to changes in the proton/neutron ratio, one of this article's 
main purposes. 

\subsection{Brief overview of findings and observations from the literature} 
\label{subsec4b}

Most theoretical studies  have been 
conducted in symmetric matter and at zero temperature. As mentioned earlier, results differ
considerably. 
In Ref.~\cite{Zhang}, three-body forces were found to induce a stronger suppression of the cross section as    
compared with Brueckner calculations with two-body forces only.       
 This effect originated from enhancement of the repulsive 
component of the effective interaction, but mostly from reduction of the density of states in the entrance
and exit channels due to the rearrangement term in the self-energy, which can also be traced back to    
the three-body force \cite{Zhang}. 

Alm {\it et al.} \cite{Alm1,Alm2} considered in-medium cross sections at finite temperature and 
observed a strong enhancement at low temperature which might be attributed to the onset of superfluidity.
Such enhancement was found to be crucially determined by the inclusion of hole-hole scattering in the 
Pauli operator. 

The predictions from Ref.~\cite{LM} are based on DBHF calculations of the (real) $R$-matrix in 
symmetric nuclear matter. 
They can be parametrized as
\begin{equation}
\sigma _{np}^{med} = (31.5 + 0.092 \; |20.2 - E_{lab}^{0.53}|^{2.9}) \frac{1.0 + 0.0034E_{lab}^{1.51} \rho^2}
{1.0 +21.55 \rho^{1.34}}  
\label{LMnp}
\end{equation}
and 
\begin{equation}
\sigma _{pp}^{med} = (23.5 + 0.00256 \; (18.2 - E_{lab}^{0.5})^{4}) \frac{1.0 + 0.1667E_{lab}^{1.05} \rho^3}
{1.0 +9.704 \rho^{1.2}} \; . 
\label{LMpp}
\end{equation}
Rather different conclusions 
were reached in Ref.~\cite{hh1}, where                 
the in-medium cross section for collision of two slabs of nuclear matter was found to increase with density.

In a phenomenological approach \cite{PP,Gale,LC,Negel} 
the NN cross sections in the medium are scaled with the                                         
factor
\begin{equation}
\frac{\sigma_{NN}^*}{\sigma_{NN}^{free}}=
\Big (\frac{\mu_{NN}^*}{\mu_{NN}} \Big)^2 \; , 
\label{scaling} 
\end{equation}
where 
$\mu_{NN}^*$ and $\mu_{NN}$ are the reduced masses of the colliding nucleon pairs in the medium and in vacuum, 
respectively.

Before we proceed, some comments are in place concerning the meaning 
of {\it isospin dependence} of the NN cross section. 
Generally, isospin dependence is understood as the mechanism by which 
the ratio $\frac{\sigma_{np}}{\sigma_{pp}}$ changes in the medium. Since                               
$\sigma_{np}$ and $\sigma_{pp}$ differ in free space due to the fact that only one isospin state is allowed in the latter,
the evolution of this ratio in the medium indicates 
to which extent 
partial waves with different isospin exhibit different behavior as a function of density. 

In our work, 
of course we have carefully considered differences between 
$\sigma_{pp}$ and 
$\sigma_{np}$ and their density dependence, but we have gone beyond that point and also addressed
differences among $\sigma_{pp}$,
$\sigma_{nn}$, and 
$\sigma_{np}$ which are 
induced by the presence of different proton and neutron densities (i.~e., isospin-asymmetric medium). 
Of course these are more subtle and, accordingly, more difficult to discern experimentally.
Nevertheless, we will demonstrate that these effects can be non-negligeable and 
do provide additional insight into how the medium separates the dynamics of 
protons and neutrons. Furthermore, they should be included when addressing 
the mean free path of a proton or a neutron in IANM.            

Experimentally, evidence has been reported for in-medium modification of NN cross sections
based on heavy ion collisions. In particular, studies of collective flow have provided
strong indication that the cross section is reduced in the medium 
 \cite{West,Xu,Daniel}.                   
The empirical relation 
\begin{equation}
\sigma _{NN}^{med} = \Big (1 + a \frac{\rho}{\rho_0} \Big )\sigma_{NN}^{free} \; , 
\label{emp}
\end{equation}
with $a \approx -0.2$, was found to be in better agreement with flow data as compared to 
calculations that made use of free-space cross sections \cite{KWB}. 
More recent studies of the stopping power and collective flow at SIS/GSI energies provided indications
that the NN in-medium cross sections are reduced at low energy but enhanced at high energy \cite{ZD07}. 

Although 
the nuclear stopping power in heavy-ion collisions has been found to be sensitive to medium effects on the 
NN cross sections, it is insufficient to discern isospin dependence \cite{LC08}.
Isospin tracers such as the neutron/proton ratio of free nucleons or the ratio of mirror nuclei \cite{DL05} have been 
proposed as potential probes of isospin dependence. 

In conclusion, the current status can be summarized as follows: considerable theoretical effort has 
been spent on the issue of the in-medium dependence of $NN$ cross sections, but much less on their isospin
dependence. From the experimental standpoint, 
there is evidence of in-medium reduction of the NN cross sections, 
but observables that can unambiguosly resolve the isospin dependence 
have not clearly been identified. A likely candidate seems to be the neutron/proton ratio of free 
nucleons, specifically at backward rapidities/angles, in reactions involving radioactive beams in inverse
kinematics. 
For instance, in Ref.~\cite{DL05} isospin sensitivity is tested by comparing rapidity distributions 
in heavy ion collisions with different assumptions for the NN cross sections, such as 
$\sigma_{np}$=$\sigma_{pp}$=$\sigma_{pp}^{free}$, or
$\sigma_{np}$=$\sigma_{np}^{free}$ and $\sigma_{pp}$=$\sigma_{pp}^{free}$, or
$\sigma_{np}$=$\sigma_{pp}$= 0.5($\sigma_{np}^{free}+\sigma_{pp}^{free}$).     
Using different 
$\sigma_{np}/ \sigma_{pp}$ ratios has impact on the transfer of neutrons or protons from 
forward to backward rapiditities, an effect which is opposite for neutrons and protons and thus reflects 
on the isospin asymmetry in a measurable way \cite{LC08}. 

\subsection{Our approach to in-medium $NN$ cross sections} 
\label{subsec4c}

  The nuclear matter calculation described in Section {\bf 3} provides, 
 along with the EoS, 
 the single-proton/neutron potentials as well as their parametrizations in 
 terms of effective masses, see Eq.~(\ref{Ui3}). Those effective masses, together with 
 the appropriate Pauli operator (depending on the type of nucleons involved), 
  are then used in a separate calculation of the in-medium       
  reaction matrix under the desired kinematical conditions.

The medium effects which we include in the calculation of the $G$-matrix, see Eq.~(\ref{Geq}), are:
Pauli blocking of the intermediate (virtual) states for two nucleons
with equal or different Fermi momenta; dispersive effects on the single-particle energies;
density-dependent nucleon spinors in the NN potential (Dirac effect). 

 Our calculation is controlled by the total density, $\rho$, and                    
 the degree of asymmetry,
$\alpha=(\rho_n - \rho _p)/(\rho_n + \rho_p)$.                                   
For the case of
identical nucleons, the $G$-matrix is calculated using the appropriate
effective mass, $m_i$, and the appropriate Pauli operator, $Q_{ii}$,                      
depending on $k_F^i$,            
where $i=p$ or $ n$. 
For non-identical nucleons, we use 
the ``asymmetric'' Pauli operator, $Q_{ij}$, depending on both 
$k_F^n$ and
$k_F^p$ \cite{AS03}. We recall that                                   
$k_F^n$ and
$k_F^p$ change with increasing neutron 
fraction according to Eqs.~(\ref{kfn}-\ref{kfp}). 

 In the usual free-space scattering scenario, the cross section is typically           
 represented as a function of the incident laboratory energy, 
 which is uniquely related to the nucleon momentum
 in the two-body c.m. frame, $q$ (also equal to one-half the 
 relative momentum of the two nucleons), through the well-known
 formula $T_{lab} = 2q^2/m$.                                       
 In nuclear matter, though, 
 the Pauli operator depends also on the total 
 momentum of the two nucleons in the nuclear matter rest frame. 
 For simplicity, here we use in-vacuum kinematics to define the total two-nucleon 
 momentum in the nuclear matter rest frame (that is, the target nucleon is, on
 the average, at rest).

Another issue to consider when addressing in-medium cross sections is the non-unitary nature of the interaction. 
In free space, and in absence of inelasticities, the (real) $R$-matrix and the (complex) $T$-matrix formalisms are
equivalent. However, 
due to the presence of Pauli blocking, which restricts the accessible spectrum of momentum states,
the in-medium  scattering matrix does not obey the free-space unitarity relations through
which phase-shift parameters are defined and from which it is customary to          
determine NN scattering observables.
Therefore, we believe that the in-medium cross section 
should be calculated from the complex $G$-matrix amplitudes (as obtained from Eq.~(21)), 
and that significant loss of information may
result if the $R$-matrix is used instead. 

A first step to obtain 
the in-medium total elastic cross section is to integrate              
the elastic differential cross section,                  
\begin{equation}
\frac{d \sigma}{d \Omega} = \frac{(m^*)^4}{4 \pi ^2 s^*}       
|{\hat G}(q,q,\theta)|^2 \; , 
\label{diff} 
\end{equation}
where ${\hat G}$ is the amplitude obtained by summing the usual 
partial wave helicity matrix elements, $m^* = m+U_S$ (see definition below Eq.~(27)), and 
$s^* = 4((m^*)^2 + q^2)$. 

\begin{figure}
\begin{center}
\vspace*{-1.0cm}
\hspace*{-2.0cm}
\scalebox{0.4}{\includegraphics{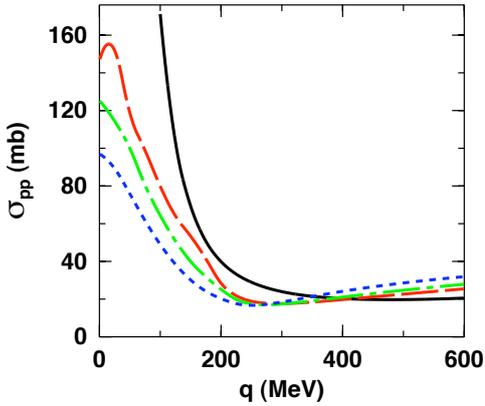}} 
\vspace*{-3.0cm}
\caption{(color online) $pp$ total elastic cross section in symmetric nuclear matter                
at various densities as   
a function of the NN relative momentum. Predictions are obtained from Eq.~(\ref{diff}) integrated
over the whole solid angle. 
The dashed (red), dash-dotted (green), and dotted (blue) curves correspond to values of the Fermi monentum 
equal to 1.1, 1.3, and 1.5 fm$^{-1}$, respectively. 
The values in free space are also shown (solid black). 
} 
\label{three}
\end{center}
\end{figure}

\begin{figure}
\begin{center}
\vspace*{-1.0cm}
\hspace*{-2.0cm}
\scalebox{0.4}{\includegraphics{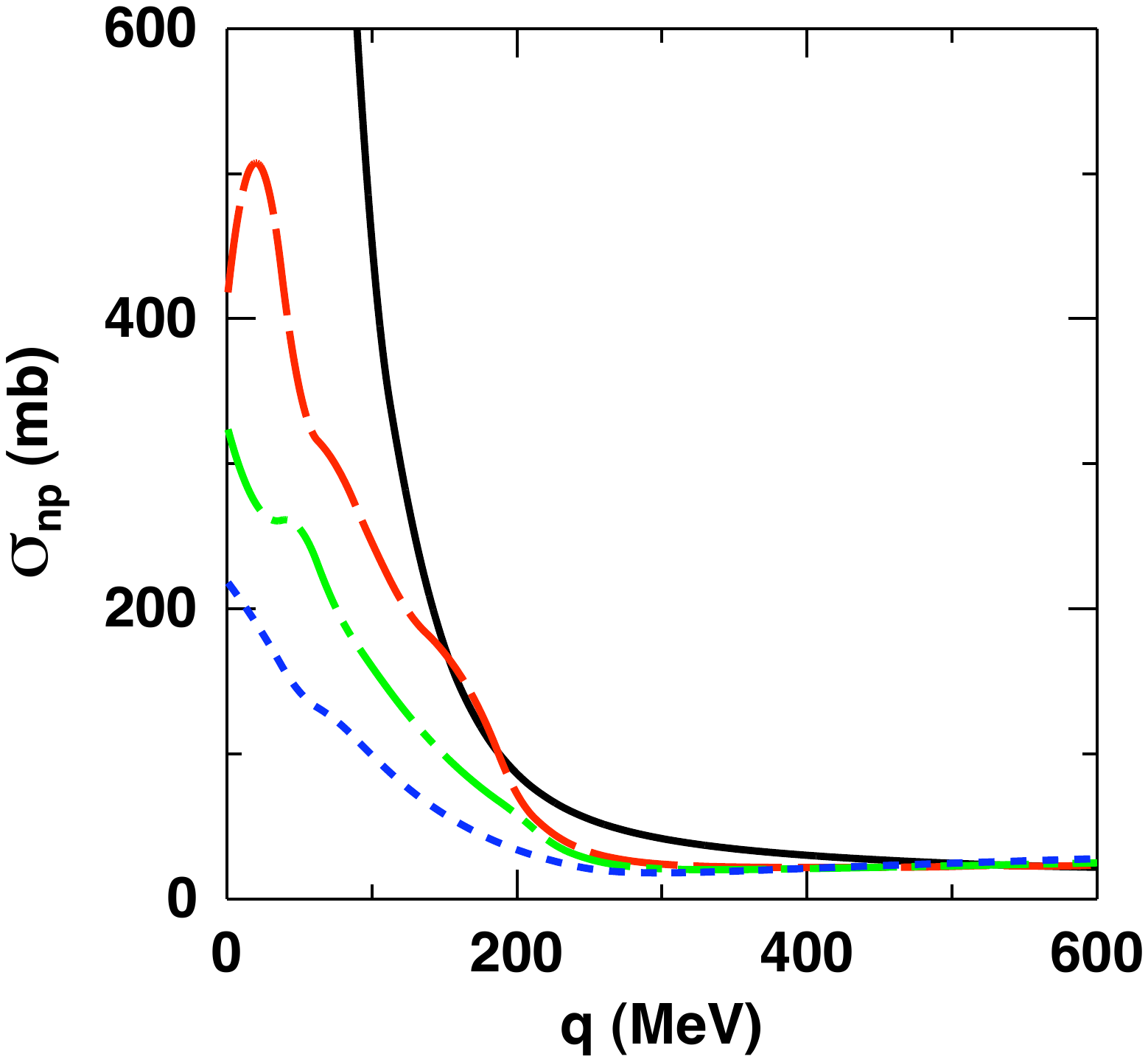}} 
\vspace*{-3.0cm}
\caption{(color online) As in Fig.~\ref{three} for $np$ scattering.                                            
} 
\label{four}
\end{center}
\end{figure}

Representative results from the procedure outlined above are shown in Figs.~\ref{three}-\ref{four}. 
There, we display $pp$ and $np$ cross sections as a function of the NN momentum in 
the center-of-mass of the pair and for three different densities of symmetric matter corresponding to Fermi momenta equal to 1.1 fm$^{-1}$, 1.3 fm$^{-1}$, and 1.5 fm$^{-1}$, respectively. The free-space predictions 
are also included. 
For fixed momentum, the cross section typically decreases with increasing
density. Likewise, for fixed density, generally it decreases as a function
of momentum.
There is a clear tendency, though, of the in-medium predictions 
to rise again with density for the higher momenta.                               
This effect   
 was already observed in previous DBHF calculations \cite{LM,Fuchs}.
We determined that it originates from the presence of the effective mass in the NN potential
and is different in nature than any of the ``conventional'' medium effects.
We found it to be particularly pronounced in isospin-1 partial waves, and thus 
relatively more important in the $pp$ channel as compared to the $np$ one, as is apparent
from a comparison of Fig.~\ref{three} and  Fig.~\ref{four}. 

At the lowest energies and for the lower densities, the cross sections show some enhancement before starting to decrease           
monotonically, a feature that is more pronounced 
in the $np$ channel (Fig.~\ref{four}), suggesting a stronger contribution from T=0 partial waves. We found that the presence and size of such enhancement is dependent on the choice adopted for $P$, the total 
momentum of the pair. 
 The structures seen in the figures are  most likely the result of competition among effects which would rise or lower the cross section. 
Notice that medium effects applied only on the $G$-matrix amplitudes but not on the phase space factor would
increase, rather than lower, the cross section. 

\begin{figure}
\begin{center}
\vspace*{1.0cm}
\hspace*{-2.0cm}
\scalebox{0.35}{\includegraphics{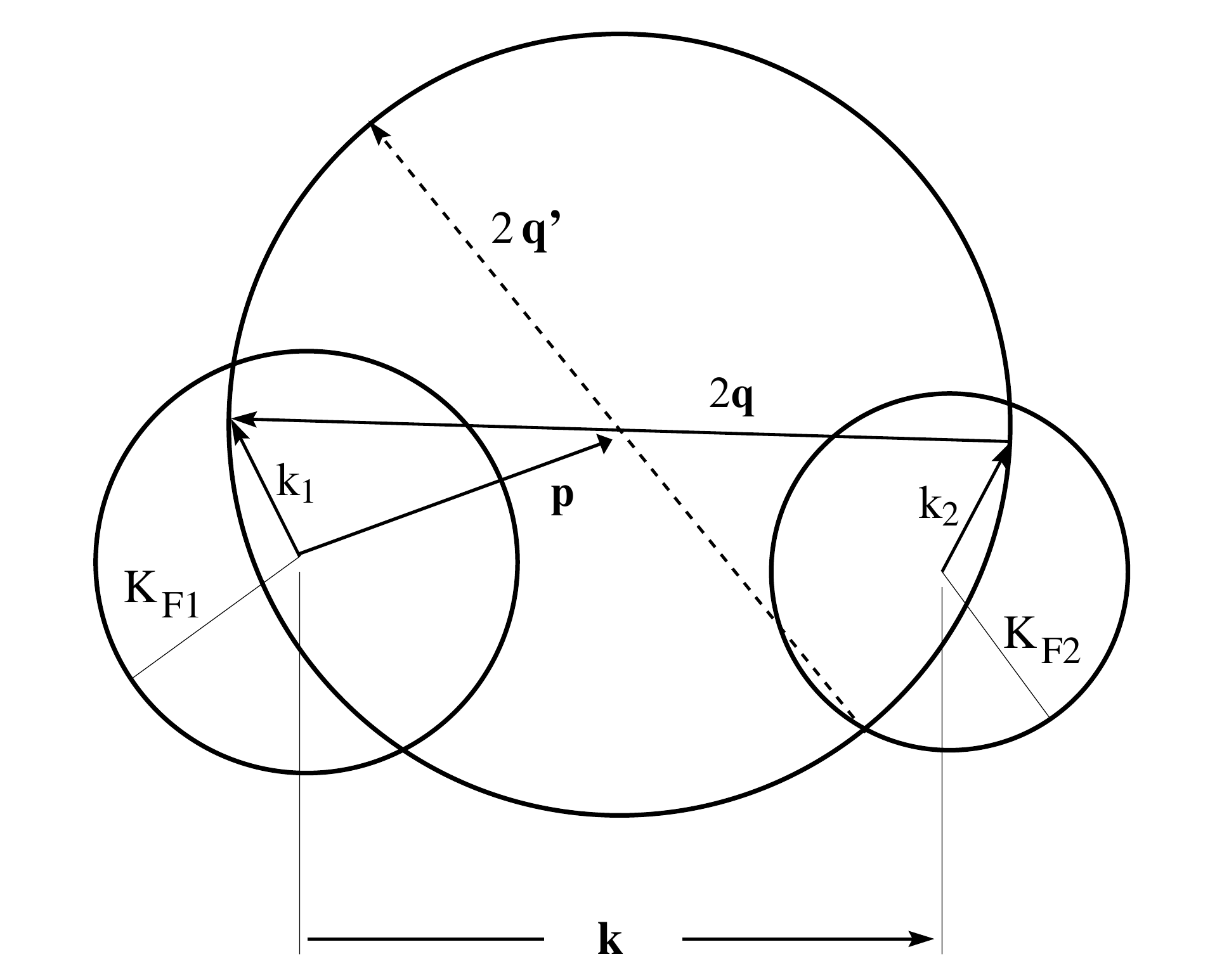}} 
\vspace*{1.0cm}
\caption{Geometrical representation of Pauli blocking in nucleus-nucleus collisions.                           
} 
\label{carlos}
\end{center}
\end{figure}

The predictions shown in Fig.~\ref{three}-\ref{four} are only a baseline as they do not yet include Pauli blocking of the final states,
an important effect for a realistic consideration of physical scattering. 
For that purpose, the allowed solid angle into which the nucleon final momenta are allowed to scatter must be restricted. 
We will discuss this aspect next. 

\begin{figure}
\begin{center}
\vspace*{-1.0cm}
\hspace*{-0.9cm}
\scalebox{0.5}{\includegraphics{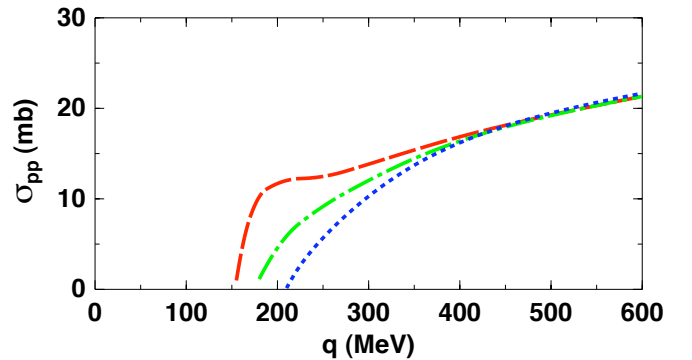}} 
\vspace*{-5.0cm}
\caption{ (color online)                    
$pp$ total effective cross section in symmetric nuclear matter                
at the same densities as in Figs.~\ref{three}-\ref{four} and as a function of the NN relative momentum.
The predictions are obtained from Eq.~(\ref{dom}).           
} 
\label{five}
\end{center}
\end{figure}

\begin{figure}
\begin{center}
\vspace*{-1.0cm}
\hspace*{-0.95cm}
\scalebox{0.5}{\includegraphics{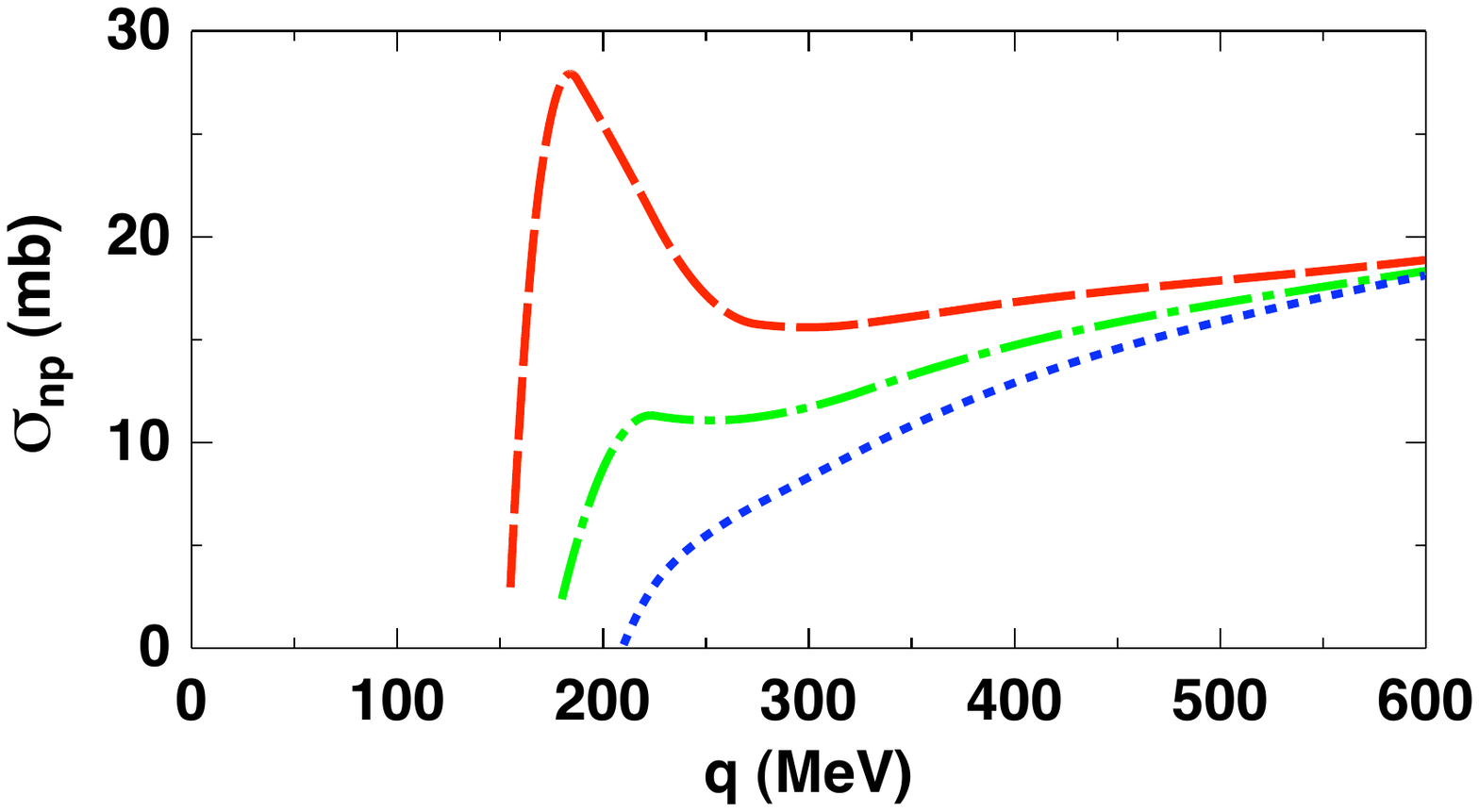}} 
\vspace*{-5.0cm}
\caption{ (color online)                    
As in Fig.~\ref{five}, for $np$ scattering.                      
} 
\label{six}
\end{center}
\end{figure}

\begin{figure}
\begin{center}
\vspace*{-1.0cm}
\hspace*{-0.95cm}
\scalebox{0.5}{\includegraphics{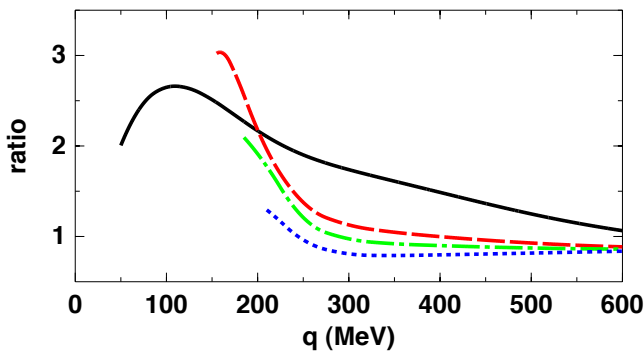}} 
\vspace*{-5.0cm}
\caption{ (color online)                    
Ratio of the $np$ and $pp$ cross sections from the previous two figures as a function of the 
momentum. The solid (black) curve shows the free-space values. 
} 
\label{extra}
\end{center}
\end{figure}

\begin{figure}
\begin{center}
\vspace*{-1.0cm}
\hspace*{-0.95cm}
\scalebox{0.5}{\includegraphics{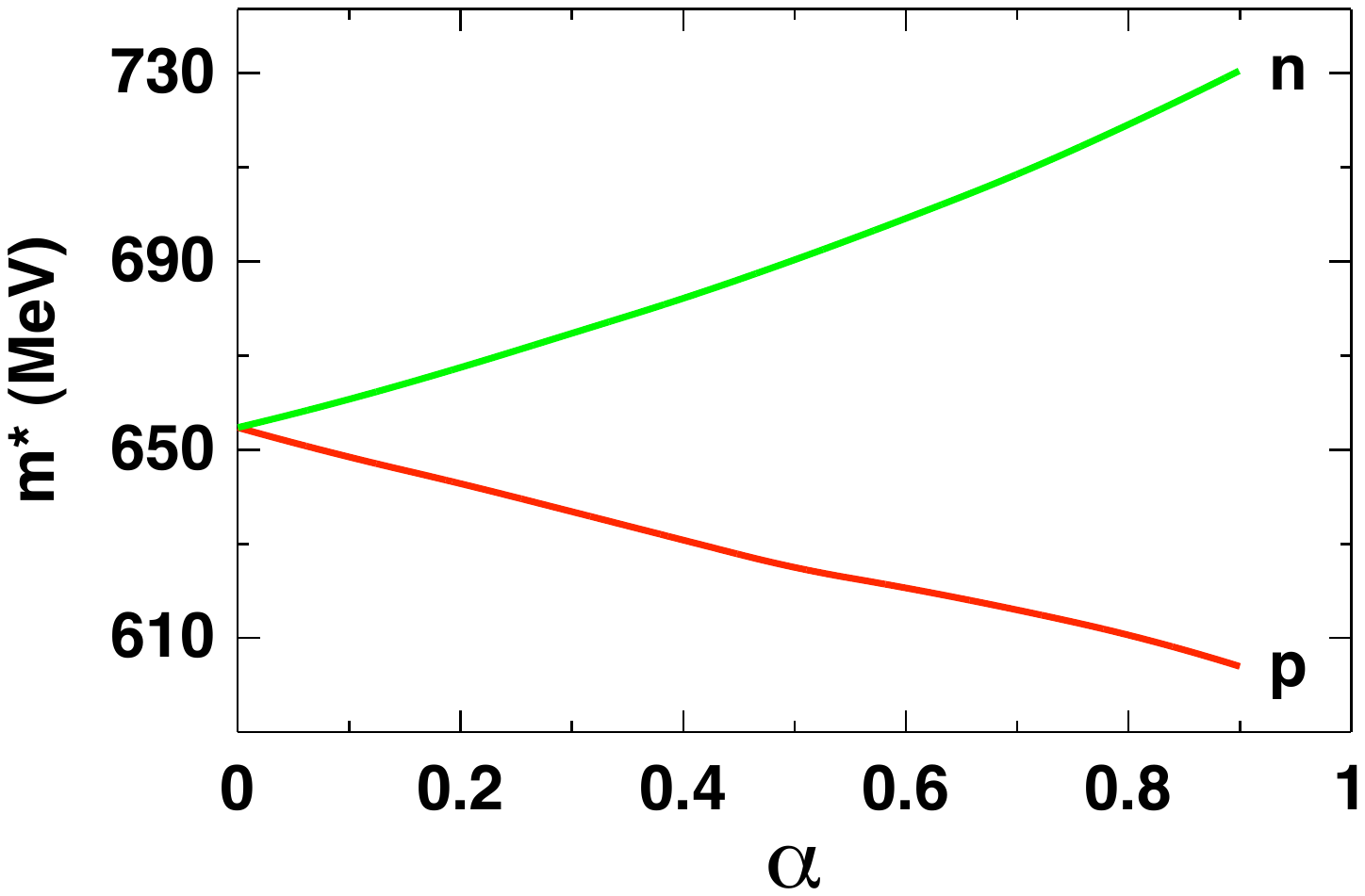}} 
\vspace*{-5.0cm}
\caption{(color online) Neutron and proton effective masses in IANM as a function of the neutron excess parameter. The 
total density is fixed and corresponds to a Fermi momentum equal to 1.3 fm$^{-1}$. 
} 
\label{seven}
\end{center}
\end{figure}

In nucleus-nucleus scattering,                                            
 two interacting {\it nucleons} within each colliding {\it nucleus} can have momenta that are, in general, 
off the symmetry axis defined by the relative momentum of the centers of the colliding nuclei, 
{\bf k}. 
To consider such case, one defines an average effective NN cross section as \cite{HRB} 
\begin{equation}
{\bar \sigma}_{NN}(k) = \frac{1}{V_{F1}V_{F2}} \int d{\bf k}_1 d{\bf k}_2 \frac{2q}{k}\sigma_{NN}(q) \int _{Pauli} d \Omega \; , 
\label{sigav}
\end{equation} 
where $k_1$ and $k_2$ (the momenta of the two nucleons relative to their respective nuclei) are smaller than $k_{F1}$ and
$k_{F2}$, respectively, and the angular integrations 
extend over all possible directions of 
${\bf k_1}$ and                                                               
${\bf k_2}$ allowed by Pauli blocking. The total and relative momenta of the two-nucleon pair               
 are given as $2{\bf q} = {\bf k_2}+{\bf k}-{\bf k_1}$, and $2{\bf P} = {\bf k_{1}} + {\bf k_{2}} + {\bf k}$ 
\cite{HRB}. 
Figure~\ref{carlos} shows the momenta appearing in Eq.~(\ref{sigav}) and the geometry of Pauli blocking.
Vectors         
 ${\bf k_{1}}$ and $ {\bf k_{2}}$ must remain smaller than the radii of their respective 
Fermi spheres, and vector 
  $2{\bf q}$ can rotate while keeping constant           
magnitude  
  (which defines the scattering sphere).                                     

 The NN cross section in the integrand of Eq.~(\ref{sigav}) corresponds to those shown in Fig.~\ref{three}-\ref{four} (or their free-space counterparts, a choice often encounered in the literature).  
$V_{F1}$ and $V_{F2}$ are the volumes of the two (in general different) Fermi spheres.                                  
A corresponding expression can be worked out which is suitable for nucleon-nucleus reactions \cite{HRB}, 
involving only one Fermi sphere.

With regard to nucleon-nucleus reactions in particular, it should also be mentioned that the             
optical model can be a powerful tool to constrain single-particle properties in nuclear matter. 
The microscopic optical model potential (OMP), which is typically obtained by folding the microscopic
$G$-matrix with the nuclear density, can be compared with the volume term of the empirical optical potential 
obtained from fits to nuclei. (Spin-orbit, surface, and Coulomb terms do not play a role in infinite nuclear 
matter.) 
Such study was reported in Ref.~\cite{Li+}, using the BHF approach together with microscopic three-body forces.
The single-proton and the 
single-neutron potentials in IANM (both elastic and absorptive parts) were compared to the volume term of the
fitted potential as a function of isospin asymmetry. Overall the agreement was found to be good, 
although the empirical absorptive part showed no clear evidence of isospin splitting, 
most likely due to isospin resolution being of the same order of magnitude as the uncertainty in the fit.
We close this short detour by noting that,
with appropriate folding of the nucleon-nucleus OMP, one can build an OMP suitable
for applications in nucleus-nucleus collisions. 

Back to our cross sections,
equation~(\ref{sigav}) is what we have applied in Ref.~\cite{CSB} in preparation for applications
to ion-ion scattering. 
On the other hand, 
here our focus is on medium and medium-induced isospin effects rather than a specific type of reaction. Thus, 
to demonstrate those effects in a more transparent way, 
we will adopt a simpler definition and      
calculate 
the effective NN cross section as                   
\begin{equation}
\sigma(q,P,\rho)= \int \frac{d\sigma}{d\Omega}                               
Q(q,P,\theta, \rho)d\Omega , 
\label{dom} 
\end{equation}
where
$\frac{d\sigma}{d\Omega} $                                    
is the differential cross section obtained from the $G$-matrix elements as in Eq.~(\ref{diff}) and 
$Q$ signifies that Pauli blocking is applied to the final configurations. 
We take 
${\bf q}={\bf P}={\bf k}/2$ and assume that the target nucleons are initially        
at rest, as in a typical free-space
scattering scenario. 
A connection with  physical scattering can be made considering 
a nucleon bound in a nucleus (or, more ideally, in nuclear matter, as in our case) through the mean field. If such 
nucleon is struck, (for instance, as in a $(e,e^{'})$ reaction), it may subsequently              
 scatter from another nucleon. This is the scattering we are describing.                            

The presence of the Pauli operator in Eq.~(\ref{dom}) restricts the integration domain
to                              
\begin{equation}
\frac{k_F^2-P^2-q^2}{2Pq} \leq  \cos \theta \leq
\frac{P^2+q^2-k_F^2}{2Pq} \; .                             
\label{cos} 
\end{equation}
The integral becomes zero if the upper limit is negative, whereas 
the full angular range is allowed if the upper limit is greater than one.
Note that the angle $\theta$ in Eq.~(\ref{cos}), namely the 
angle between the directions of $\vec{q}'$ (the relative momentum after scattering) and $\vec{P}$, is also the colatitude
of $\vec{q}'$ in a system where the $z$-axis is along 
the (conserved) vector $\vec{P}$ and, thus, it coincides with the scattering angle to be 
integrated over in Eq.~(\ref{dom}). 
            
Ignoring Pauli blocking on the final momenta amounts to setting $Q$=1 in the integrand above, 
as done in previous works \cite{LM,Fuchs}, and results in predictions such as those shown in Fig.~\ref{three}-\ref{four}.
Notice that the cross sections displayed in Fig.~\ref{three}-\ref{four} and those we will obtain from Eq.~(\ref{dom}) can be 
dramatically different, nor should one expect agreement, as the restriction Eq.~(\ref{cos}) can completely 
suppress the cross section is some regions of the density-momentum phase space. 
This mechanism should be  
included for a realistic calculation of the nucleon mean free path in nuclear 
matter, which must approach large values as the scattering probabilty goes to zero.

\subsection{Effective cross sections in       
symmetric and asymmetric matter} 
\label{subsec4d}

The effective cross sections shown in this Section are obtained from Eq.~(\ref{dom}). 
We begin with representative results in symmetric nuclear matter (SNM). 
For this purpose, we need only to address $pp$ and $np$ cross sections, whereas in 
IANM we will also need to distinguish between the $pp$ and the $nn$ cases, as anticipated 
in Section~{\bf 4.2}. 

\noindent
\underline{{\bf $pp$ (or $nn$) cross sections in SNM:}}\\    
These are
shown in Fig.~\ref{five} as a 
function of $q$ at selected densities.                         
The given range of $q$ corresponds to values of the                              
in-vacuum laboratory kinetic energy up to                                         
approximately 800 MeV. 
(In passing, we recall that a good quality OBE potential is able to 
describe NN elastic scattering up to nearly 1000 MeV. Thus, as long as one is not concerned
with pion production reactions, it should be reasonable to calculate the in-medium cross section
from the elastic part of the NN interaction as described by the OBE model.) 
The densities associated with the chosen Fermi momenta are equal to 0.067 fm$^{-3}$,    
0.148 fm$^{-3}$, and 
0.228 fm$^{-3}$, respectively. 
Due to the presence of the Pauli operator in Eq.~(\ref{dom}), it is apparent that the 
cross sections will become identically zero at certain densities depending on the 
value of the momentum. Thus, cross sections calculated with this mechanism 
can be quite
different, both quantitatively and qualitatively, than those shown in      
 Fig.~\ref{three}.

Naturally, at the lower momenta there is 
strong sensitivity to any small variation of the Fermi momentum as one is           
approaching the region where the cross
section vanishes.                        

At the densities considered in the figure, 
the cross sections mostly grow with                    
energy. This is to be expected and due to the fact that the Pauli operator in Eq.~(\ref{dom}) becomes less effective at the 
higher energies. 
The underlying energy dependence displayed in Fig.~\ref{three}, combined with the 
trend of the cross section (as given in Eq.~(\ref{dom})) to increase with energy due to reduced Pauli blocking, results into a broad 
local maximum which disappears as density increases. 

Other mechanisms responsible for 
the tendency of the cross section 
to rise at the higher momenta were already pointed out in conjunction with Fig.~\ref{three}, and play the 
same role in Fig.~\ref{six}. 

\noindent
\underline{{\bf $np$ cross sections in SNM:}}\\       
These are
shown in Fig.~\ref{six}.                            
Similar comments apply as those made above for the $pp$ case.         
There is                                 
very strong $k_F$-sensitivity at the lowest momenta.                              
As discussed above for the $pp$ case,                       
the cross section first grows with                    
energy due to the fact that the Pauli operator in Eq.~(\ref{dom}) becomes less effective the higher the 
energy. 
The $np$ cross section, though, shows a much more pronounced peak structure (as compared to the $pp$ one)
at relatively low densities. The peak is washed out as density increases. As discussed previously, these 
structures are the result of the energy dependence shown in 
Fig.~\ref{four} and the Pauli operator in Eq.~(\ref{dom}) cutting less of the solid angle at higher momenta. 

\noindent 
\underline{{\bf Summary of $pp$ {\it vs.} $np$ in SNM.}}\\       
The comparison of 
Fig.~\ref{five} with Fig.~\ref{six} reveals the isospin dependence as 
discussed in Section {\bf 4.2}. Such comparison shows that the relation between 
$\sigma_{pp}$ and 
$\sigma_{np}$ can be altered considerably as a function of density as compared to 
the vacuum, because 
the two cross sections exhibit different structures in the medium, 
particularly at specific densities and momenta.                                               
This indicates non-trivial isospin dependence                                
in the way partial waves are impacted by the medium as a function of momentum, as demonstrated in 
in Fig.~\ref{extra}. There we display the ratio of the cross 
sections from Fig.~\ref{six} and Fig.~\ref{five},                                                           
$\frac{\sigma_{np}}{\sigma_{pp}}$,               
along with its value in free space. The behavior of this 
ratio as a function of energy and density reflects the previously discussed trend of the $np$ cross section to 
be enhanced in the medium at the lower momenta as well as the tendency of the $pp$ cross section to grow more rapidly 
at high momenta. 

\noindent 
\underline{{\bf $pp$ and $nn$ cross sections in IANM.}}\\       
In IANM,  as the neutron population increases, the single-neutron and the 
single-proton potentials, Eqs.~(22-23), become more repulsive and more attractive, respectively. 
Figure \ref{seven} shows how the corresponding                                                         
 effective masses change as a function of the neutron excess
parameter for fixed total density As the cross section depends strongly on the effective mass, this can be insightful when interpreting the predictions.              

\begin{figure}
\begin{center}
\vspace*{-1.0cm}
\hspace*{-1.8cm}
\scalebox{0.5}{\includegraphics{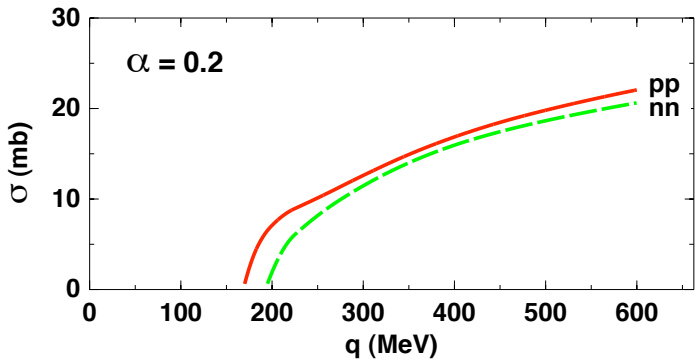}} 
\vspace*{-5.0cm}
\caption{ 
$pp$ and $nn$ total effective cross sections in IANM versus the NN  relative momentum for fixed total density and asymmetry.   
$k_F$=1.3 fm$^{-1}$.                        
} 
\label{nine}
\end{center}
\end{figure}

\begin{figure}
\begin{center}
\vspace*{-1.0cm}
\hspace*{-1.5cm}
\scalebox{0.5}{\includegraphics{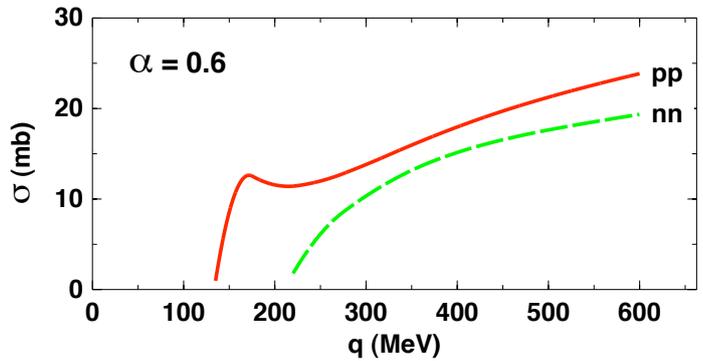}} 
\vspace*{-5.0cm}
\caption{ 
As in Fig.~\ref{nine} for                                                                                         
a larger value of $\alpha$.         
} 
\label{ten}
\end{center}
\end{figure}

Predictions for $pp$ and $nn$ total cross sections in isospin-asymmetric
matter are                                                                                   
shown in Figs.~\ref{nine}-\ref{ten}, as a function of $q$ and fixed total density and degree of asymmetry.
A smaller and a larger degree of asymmetry are considered in Fig.~\ref{nine} and \ref{ten}, respectively.
Since $pp$ ($nn$) scattering is only impacted by the proton (neutron) 
Fermi momentum, these cross sections are calculated using the proton (or neutron)
Fermi momentum in the Pauli operator that would be appropriate for symmetric matter at that density. 

Concerning the  momentum dependence, similar comments apply as those
made with regard to Fig.~\ref{five}. Note that for $\alpha > 0$, the  
the $nn$ cross section is (almost always) smaller than the $pp$ cross section.
This is due to the                                                              
additional Pauli blocking included in Eq.(\ref{dom}) and the fact that the 
neutron's 
Fermi momentum is larger than the proton's. Therefore,                         
for the same total nucleon density,            
$\sigma_{nn}$ may be entirely suppressed at momenta where $\sigma_{pp}$ is 
still considerably larger than zero. (See, in particular, Fig.~\ref{ten}.)

\begin{figure}
\begin{center}
\vspace*{-1.0cm}
\hspace*{-0.9cm}
\scalebox{0.5}{\includegraphics{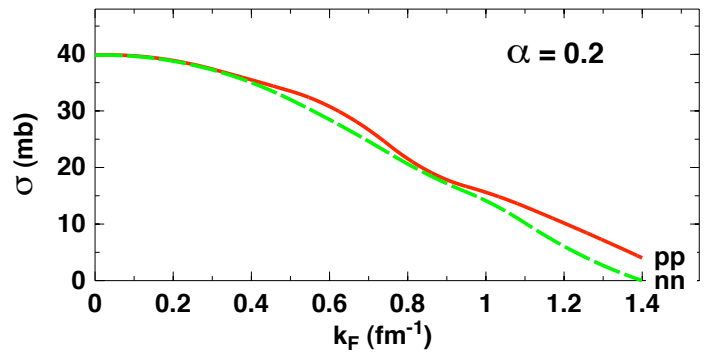}} 
\vspace*{-5.0cm}
\caption{Density dependence of the 
$pp$ and $nn$ total effective cross sections in IANM at fixed momentum ($q$=200 MeV) and asymmetry.                                  
} 
\label{eleven}
\end{center}
\end{figure}

\begin{figure}
\begin{center}
\vspace*{-1.0cm}
\hspace*{-0.2cm}
\scalebox{0.5}{\includegraphics{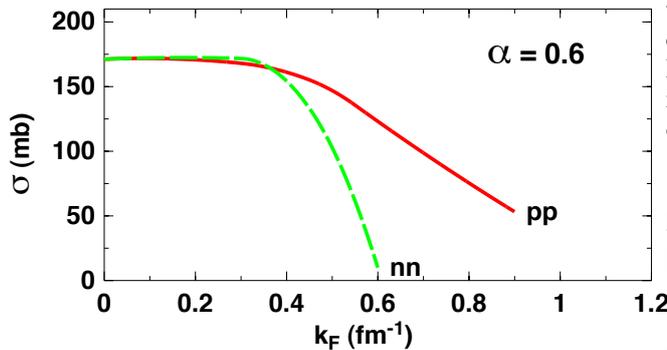}} 
\vspace*{-4.9cm}
\caption{ 
As in Fig.~\ref{eleven} for                                                                                         
$\alpha$=0.6 and $q$=100 MeV. 
} 
\label{twelve}
\end{center}
\end{figure}

Additional results are displayed
in Fig.~\ref{eleven}-\ref{twelve}. There, we show some interesting trends of the $pp$ and $nn$ cross sections 
{\it versus} density for fixed momentum and asymmetry. 
Although at low density the $nn$ and $pp$ cross sections are nearly equal, 
Pauli blocking soon takes over and clearly separates the $pp$ and $nn$ cases. 
Particularly for large values of $\alpha$, the $pp$ cross section
``survives'' larger densities than the $nn$ one, due to the fact that  
the proton Fermi momentum, $k_F^p$=$k_F(1-\alpha)^{1/3}$,              
is much smaller than the neutron one for large $\alpha$.
The splitting is remarkable at low momenta and for high degree of 
isospin asymmetry, compare Fig.~\ref{eleven} and Fig.~\ref{twelve}. 

\noindent 
\underline{{\bf $np$ cross sections in IANM.}}\\       
The effective masses of neutron and proton change with  
increasing degree of asymmetry as shown in 
Fig.~\ref{seven}, which clearly impacts the cross section in opposite ways (relative 
to its value in SNM at the same total density). 
At the same time, the neutron and proton Fermi momenta become larger and smaller,
respectively, which also impacts the cross section in opposite directions, through Pauli blocking. 
More precisely, the neutron increasing mass and larger Fermi momentum compete with the proton 
decreasing mass and lower Fermi momentum. 
We observed that, 
due to these competing effects, the overall 
$\alpha$ dependence of the $np$ cross section  
is very weak and can be ignored. In other words, we find that the $np$ cross section in IANM can be 
calculated as in SNM using the average Fermi momentum and the average of the neutron 
and proton effective masses. 

\noindent 
\underline{{\bf Summary of observations.}}\\       
We have discussed two levels of isospin dependence of the in-medium NN cross sections. One
concerns nucleon pairs with different total isospin, whereas the other refers to pairs of 
identical nucleons with different $z$-component of the isospin.
(The explicit $\alpha$ dependence of      
$\sigma_{np}$ was found to be very weak and ignored.) 

Interesting differences exist between 
the momentum and density dependence of 
$\sigma_{pp}$ and 
$\sigma_{np}$. These can already be seen when comparing Fig.~\ref{three} and Fig.~\ref{four},
but they become more pronounced with the inclusion of the additional Pauli blocking in the final states, 
 see 
 Fig.~\ref{five} and Fig.~\ref{six}.

With regard to identical nucleons, 
the region of the density/momentum phase space       
where $\sigma_{nn}$ is nearly or entirely suppressed whereas $\sigma_{pp}$ is 
still considerably different from zero should be a suitable ground to 
look for a signature of their difference. Figures~\ref{ten} and \ref{twelve},
compared with Fig.~\ref{nine} and Fig.~\ref{eleven}, respectively, suggest that reactions involving low momenta and
medium to high densities, together with a high degree of isospin asymmetry in the collision region, has the potential to clearly separate $pp$ and
$nn$ scattering and to 
discriminate between models which do or do not
distinguish amongst different nucleon pairs.                                                              
Recalling the comments made at the end of Section {\bf 4.2}, 
isospin dependence is expected to impact neutron and proton rapidities in opposite ways. This will 
increase the isospin asymmetry, $\alpha$, and, in turn, may generate new isospin dependence, including 
the one demonstrated in Figs.~\ref{nine}-\ref{twelve} (namely, the one which refers to pairs of identical nucleons with different
values of $T_z$). 

Naturally, isospin-sensitive description of reactions requires accurate knowledge of proton and neutron densities in the target 
and projectile, so that the appropriate cross section, $\sigma_{ij}$, can be applied at each specific
point in space where the nuclei have local baryon densities $\rho_i$ and $\rho_j$ ($i,j=n,p$). 
Determining neutron densities (through measurements of the neutron r.m.s. radius and skin) is
part of the many coherent efforts presently going on to constrain the symmetry energy and related
observables. 
Thus we close this section underlining 
the importance of {\it both} empirical constraints and microscopic 
calculations (which have true predictive power) towards a better understanding of neutron-rich
systems. Microscopic            
in-medium NN cross sections can play an important role in such endeavor. 

\section{Mean-free path of protons and neutrons in IANM}
We present here a brief discussion of the mean free path 
of nucleons in nuclear matter. 

A simple and intuitive way to define the mean free path (MFP) in terms 
of the effective cross sections discussed in the previous section is 
\begin{equation}
\lambda_p = \frac{1}{\rho_p \sigma_{pp} + \rho_n \sigma_{pn}} \; , 
\label{mfp1}
\end{equation}
with an    
 analogous definition for the neutron,               
\begin{equation}
\lambda_n = \frac{1}{\rho_n \sigma_{nn} + \rho_p \sigma_{np}} \; . 
\label{mfp2}
\end{equation}
Notice that 
the above expression can be easily interpreted as the length of the unit volume in a phase space 
defined by the effective scattering area (the cross section) and the number 
of particles/volume \cite{PP}. 

\begin{figure}
\begin{center}
\vspace*{-1.0cm}
\hspace*{-1.0cm}
\scalebox{0.5}{\includegraphics{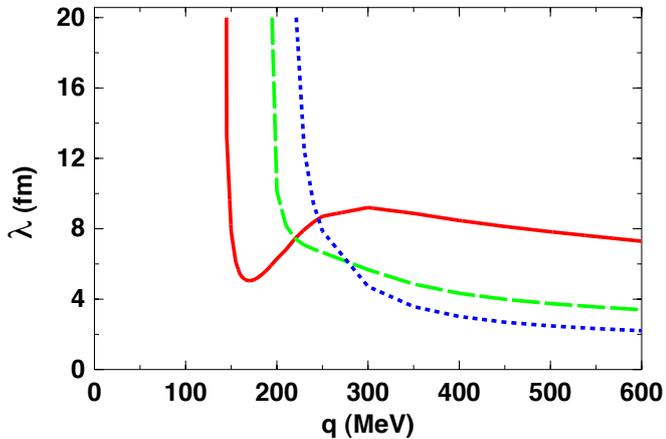}} 
\vspace*{-5.0cm}
\caption{The mean free path of a nucleon as from Eq.~(\ref{mfp1}) in symmetric matter            
as a function of momentum for three different densities corresponding to Fermi momenta equal
 1.0 fm$^{-1}$ (solid red), 
 1.3 fm$^{-1}$ (dashed green), and 
 1.5 fm$^{-1}$ (dotted blue).         
} 
\label{thirteen}
\end{center}
\end{figure}

We begin with discussing the MFP in symmetric matter, in which case
$\lambda _p$=
$\lambda _n$. This is shown in Fig.~\ref{thirteen}, as a function of the relative momentum 
and for different densities. Obviously, the mean free path approaches infinity when both cross sections
in the denominator tend to zero.                                                              
The higher the density, the higher the energy at which the MFP begins to drop rapidly. 
The lowest energy for which $\lambda$ is finite corresponds to the lowest momentum allowed 
by Pauli blocking of the final states. 
       
At the lowest density shown in the figure, the MFP displays 
some fluctuation. This corresponds to similar fluctuations in the cross sections, which we discussed 
earlier in terms of 
 competing mechanisms.                                 
Otherwise, the MFP decreases monotonically.
There is considerable density dependence.

\begin{figure}
\begin{center}
\vspace*{-1.0cm}
\hspace*{-1.0cm}
\scalebox{0.5}{\includegraphics{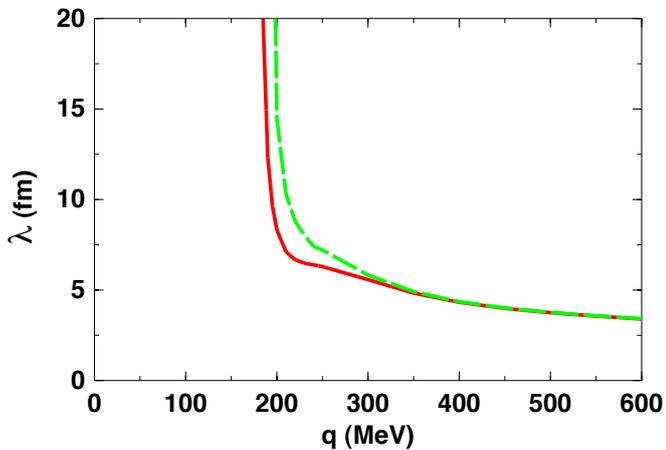}} 
\vspace*{-5.0cm}
\caption{The mean free path of a proton (solid red) and a neutron (dashed green) obtained from Eqs.~(\ref{mfp1}-\ref{mfp2}) in asymmetric matter at 
fixed density ($k_F$=1.3 fm$^{-1}$) and neutron excess parameter ($\alpha$=0.2.) 
} 
\label{fourteen}  
\end{center}
\end{figure}

\begin{figure}
\begin{center}
\vspace*{-1.0cm}
\hspace*{-1.0cm}
\scalebox{0.5}{\includegraphics{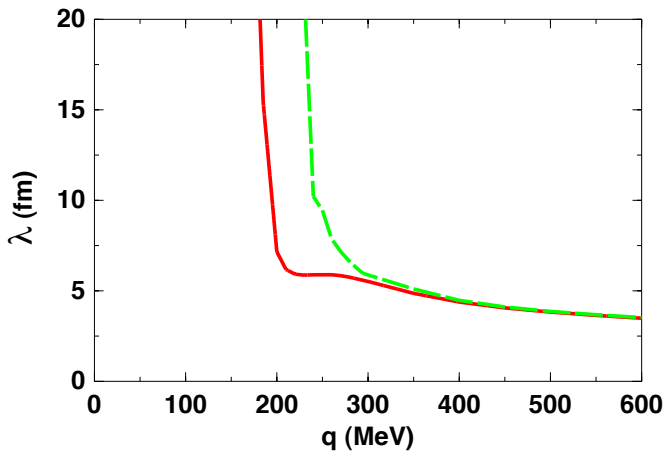}} 
\vspace*{-5.0cm}
\caption{As in the previous figure     
but for $\alpha$=0.6.  
} 
\label{fifteen}  
\end{center}
\end{figure}

Figures~\ref{fourteen}-\ref{fifteen} show the proton (solid red line) and neutron  (dashed green) MFP in asymmetric matter for fixed
total density and a low and high degree of asymmetry, respectively. 
Due to the higher neutron Fermi momentum, the $nn$ cross section is more strongly 
Pauli-blocked at the lower momenta (and thus $\lambda_n \rightarrow \infty$). 
As to be expected, $\lambda_n \approx \lambda_p$ at the higher momenta, where (all) medium
effects tend to become less important. 
Qualitatively, similar tendencies are observed in                                                                   
Figs.~\ref{fourteen} and \ref{fifteen},                                                                     
although more pronounced in the latter.                         
Overall, 
there are significant differences between a neutron and a proton MFP, which are almost entirely
due to the differences between $\sigma_{nn}$ and $\sigma_{pp}$.

\begin{figure}
\begin{center}
\vspace*{-1.0cm}
\hspace*{-1.0cm}
\scalebox{0.5}{\includegraphics{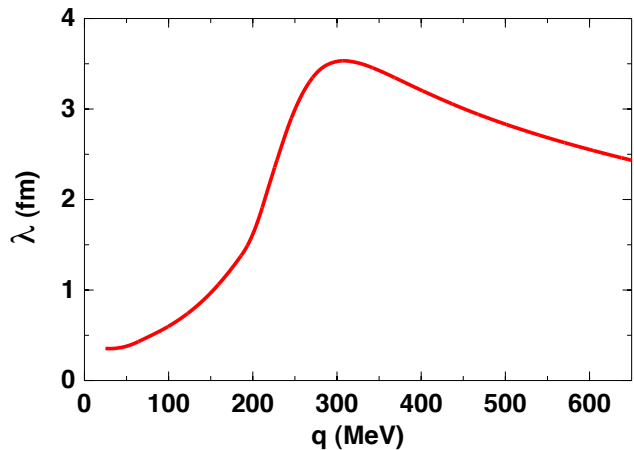}} 
\vspace*{-5.0cm}
\caption{The mean free path of a nucleon in symmetric matter at 
fixed density calculated without considerations of Pauli blocking of the final states.
The Fermi momentum is equal to 1.3 fm$^{-1}$. 
See text for more details. 
} 
\label{sixteen}  
\end{center}
\end{figure}

Reconnecting with 
the previous discussion which followed Eq.~(\ref{dom}), we also show, see Fig.~\ref{sixteen}, 
a sample of MFP calculations without considerations of Pauli blocking of the final states. 
In this case, $\lambda$ becomes very small at low incident energies, due to the large values of the 
cross section, which approaches the free-space result, rather than zero, in that region.

\section{Conclusions}
                                  
We reviewed our microscopic approach to the development of the EoS of IANM 
and, self-consistently,                   
the effective interaction in the isospin-asymmetric medium. 
Within the DBHF method, the interactions of nucleons with the medium are espressed
as self-energy corrections to the nucleon propagator. That is, the nucleons are regarded
as``dressed" quasiparticles. Relativistic effects lead to an intrinsically deonsity dependent
interaction which is approximately consistent with the contribution from the three-body force 
arising from virtual pair excitations.

The focal point has been         
how the presence of a dense hadronic medium impacts the scattering 
amplitude and thus the cross section, with particular attention to the case where 
neutron and proton concentrations are different. 
To that end, we presented microscopic calculations of total elastic cross sections for scattering of 
nucleons in symmetric and neutron-rich matter. 
Our predictions include all ``conventional'' medium effects as well as 
those associated with the nucleon Dirac wavefunction.
Pauli blocking of the final states is included in the integration of the 
differential cross section.

One of the mechanisms driving the in-medium cross sections are the neutron and proton effective
masses. In turn, these are determined by the potentials experienced by the neutron and proton
in asymmetric matter, which are part of the calculation leading to the EoS of IANM. 
Thus, in our philosophy, medium effects originating from the equation of state are consistently
incorporated in the mean field and the NN cross sections.

First, we discussed 
the basic density/momentum dependence of $pp$ and $np$ cross sections in symmetric matter.
Although they generally exhibit qualitatively similar behavior with changing energy and density,
$pp$ and $np$ effective cross
sections show some interesting differences in specific regions of the phase space.
This gives rise to isospin dependence. 

With regard to identical nucleons,
the sensitivity to the asymmetry in neutron and proton concentrations comes in through    
the                                
combined effect of Pauli blocking and changing effective masses. 
The lowering(rising) of the proton(neutron) Fermi momentum and the reduced(increased)
proton(neutron) effective mass tend to move the cross section in opposite 
directions. With Pauli blocking applied to intermediate and final
states, the final balance is that the $nn$ effective cross section
is more strongly suppressed. 

In summary,          
sensitivity to the asymmetry is non-negligible for scattering of different pairs 
of identical nucleons, and clearly separates $pp$ and $nn$ scatterings. 
The degree of sensitivity depends on the 
region of the energy-density-asymmetry phase space under consideration.

We also considered 
the mean free path of a nucleon and determined that it is affected in a significant way by  
the presence of isospin asymmetry in the medium.                          

In-medium two-body collisions are only part of the input needed for reaction calculations.
Therefore, 
in closing, we reiterate the importance of coherent effort from theory and experiment 
as well as the importance of calculations with predictive power towards improved understanding
of neutron-rich systems, both reactions and structure. 

Finally, we cannot stress enough that the behavior of microscopic in-medium cross sections can be
rather complex being the result of several, often competing, mechanisms. Therefore, microscopic 
predictions do not appear to validate a simple phenomenological {\it ansatz}, such as the effective mass scaling model in
Eq.~(\ref{scaling}).

\begin{center}
{\bf Acknowledgments}                                                           
\end{center}
Support from the U.S. Department of Energy under Grant No. DE-FG02-03ER41270 is 
acknowledged.                                                                          
I am grateful to C.A. Bertulani for insightful discussions.


\begin{thebibliography}{99} 
\bibitem{Mac89}R. Machleidt, Adv. Nucl. Phys. {\bf 19}, (1989) 189. 
\bibitem{Oya98} K. Oyamatsu, I. Tanihata, Y. Sugahara, K. Sumiyoshi, and 
H. Toki, Nucl. Phys. {\bf A634}, (1998) 3.
\bibitem{Furn} R.J. Furnstahl, Nucl. Phys. {\bf A706}, (2002) 85.
\bibitem{SL09} F. Sammarruca and P. Liu, Phys. Rev. C {\bf 79}, (2009) 057301.
\bibitem{Steiner} A.W. Steiner, Phys. Rev. C {\bf 74}, (2006) 045808, and references 
therein. 
\bibitem{Tsang+} M.B. Tsang {\it et al.}, Phys. Rev. C {\bf 86}, (2012) 015803.
\bibitem{GDR} C. Simenel, Ph. Chomaz, and G. de France, Phys. Rev. C {\bf 76}, (2007) 024609.
\bibitem{BA05} B.A. Li and L.W. Chen, Phys. Rev. C {\bf 72}, (2005) 064611.            
\bibitem{Ko09}C.M. Ko, Proceedings of the ``International Workshop on Nuclear Dynamics
in Heavy-Ion Reactions and the Symmetry Energy", Shanghai, China, August 23-25, 2009, 
Special Issue of Int. J. Mod. Phys. E {\bf 19}, (2010) 1763.
\bibitem{Lane} A.M. Lane, Nucl. Phys. {\bf 35}, (1962) 676.
\bibitem{chi} R. Machleidt and D.R. Entem, Phys. Rep. {\bf 503}, (2011) 1, and references therein. 
\bibitem{pot1} R. Machleidt,  Phys. Rev. C {\bf 63}, (2001) 024001.               
\bibitem{pot2} V.G.J. Stoks {\it et al.},  Phys. Rev. C {\bf 49}, (1994) 2950.          
\bibitem{pot3} R.B. Wiringa {\it et al.}, Phys. Rev. C {\bf 51}, (1995) 38.              
\bibitem{Thom} R.H. Thompson,  Phys. Rev. D {\bf 1}, (1970) 110. 
\bibitem{BS} E.E. Salpeter and H.A. Bethe,  Phys. Rev. {\bf 84}, (1951) 1232. 
\bibitem{GB87} G.E. Brown {\it et al.}, Comments Nucl. Part. Phys. {\bf 17}, (1987) 39.
\bibitem{AS03} D. Alonso and F. Sammarruca, Phys. Rev. C {\bf 67}, (2003) 054301.
\bibitem{FS10} F. Sammarruca, Int. J. Mod. Phys. E {\bf 19}, (2010) 1259.
\bibitem{BM84} R. Brockmann and R. Machleidt,  Phys. Lett. {\bf 149B}, (1984) 283; Phys. Rev. C {\bf 42}, (1990) 1965. 
\bibitem{HT70} M.I. Haftel and F. Tabakin,  Nucl. Phys. {\bf A158}, (1970) 1. 
\bibitem{BBU1} G.F. Bertsch and S. Das Gupta, Phys. Rep. {\bf 160}, (1988) 189. 
\bibitem{BBU2} W. Cassing, W. Metag, U. Mosel, and K. Niita, Phys. Rep. {\bf 188}, (1990) 363.
\bibitem{PP} V.R. Pandharipande and S.C. Pieper, Phys. Rev. C {\bf 45}, (1992) 791.
\bibitem{Gale} D. Persram and C. Gale, Phys. Rev. C {\bf 65}, (2002) 064611.
\bibitem{LC} B.-A. Li and L.-W. Chen, Phys. Rev. C {\bf 72}, (2005) 064611.
\bibitem{LM} G.Q. Li and R. Machleidt, Phys. Rev. C {\bf 48}, (1993) 1702;
 {\bf 49}, (1994) 566.
\bibitem{Fuchs} C. Fuchs, A. Faessler, and M. El-Shabshiry, Phys. Rev. C {\bf 64}, (2001) 024003.
\bibitem{SSRL} H.-J. Schulze, A. Schnell, G. R{\" o}pke, and U. Lombardo, Phys. Rev. C {\bf 55}, (1997) 3006.
\bibitem{Zhang} H.F. Zhang, Z.H. Li, U. Lombardo, P.Y. Luo, F. Sammarruca, and W. Zuo, Phys. Rev. C {\bf 76}, (2007) 054001.
\bibitem{Glauber} R.J. Glauber, {\it Lectures on Theoretical Physics} Vol.~I (Interscience, New York, 1959).
\bibitem{Alm1} T. Alm, G. R{\" o}pke, and M. Schmidt, Phys. Rev. C {\bf 50}, (1994) 31.
\bibitem{Alm2} T. Alm, G. R{\" o}pke, W. Bauer, F. Daffin, and M. Schmidt, Nucl. Phys. {\bf A587}, (1995) 815.
\bibitem{hh1} A. Bohnet, N. Ohtsuka, J. Aichelin, R. Linden, and A. Faessler, 
Nucl. Phys. {\bf A494}, (1989) 349.
\bibitem{Negel} J.W. Negele and K. Yazaki, Phys. Rev. Lett. {\bf 62}, (1981) 71.
\bibitem{Xu} H.M. Xu, Phys. Rev. Lett. {\bf 67}, 2769 (1991); Phys. Rev. C {\bf 46}, (1992) R389.
\bibitem{Daniel} P. Danielewicz, Acta Physica Polon. B {\bf 33}, (2002) 45.
\bibitem{West} G.D. Westfall {\it et al.}, Phys. Rev. Lett. {\bf 71}, (1993) 1986.
\bibitem{KWB} D. Klakow, G. Welke, and W. Bauer,Phys. Rev. C {\bf 48}, (1993) 1982.    
\bibitem{ZD07} Y. Zhang, Z.X. Li, and P. Danielewicz, Phys. Rev. C {\bf 75}, (2007) 034615.  
\bibitem{DL05} B.A. Li, P. Danielewicz, W.G. Lynch, Phys. Rev. C {\bf 71}, (2005) 054603.
\bibitem{LC08} Bao-An Li, Lie-Wen Chen, and Che Ming Ko, Phys. Rep. {\bf 464}, (2008) 113.
\bibitem{HRB} M.S. Hussein, R.A. Rego, and C.A. Bertulani, Phys. Rep. {\bf 201}, (1991) 279.
\bibitem{Li+} L.L. Li {\it et al.}, Phys. Rev. C {\bf 80}, (2009) 064607. 
\bibitem{CSB} B. Chen, F. Sammarruca, and C.A. Bertulani, Phys. Rev. C {\bf 87}, (2013) 054616.


\end{thebibliography}
\end{document}